\documentclass[iop]{emulateapj}

\usepackage{multirow}

\newcommand{\amp} {AMP}

\newcommand{\xmm} {{\it XMM-Newton}}
\newcommand{\chandra} {{\it Chandra}}
\newcommand{\nustar} {{\it NuSTAR}}

\newcommand{\swiftbat} {{\it Swift}/BAT}

\newcommand{\cmsq} {cm$^{-2}$}
\newcommand{\nh} {$N_{\rm{H}}$}
\newcommand{\lx} {$L_{\rm{X}}$}
\newcommand{\fx} {$F_{\rm{X}}$}
\newcommand{\chisq} {$\chi^2$}

\newcommand{\mic}{{${\mu}$m}}

\newcommand{\nev}{{\rm{[Ne\,\sc{v}]}}}

\newcommand{\degree}{{$^\circ$}}

\newcommand{\ergs}{\mbox{\thinspace erg\thinspace s$^{-1}$}}
\newcommand{\kms}{\mbox{\thinspace km\thinspace s$^{-1}$}}
\newcommand{\ergcms}{\mbox{\thinspace erg\thinspace cm$^{-2}$\thinspace s$^{-1}$}}

\newcommand{\lbol} {$L_{\rm Bol}$}
\newcommand{\mbh} {$M_{\rm BH}$}

\newcommand{\lamedd} {$\lambda_{\rm Edd}$}

\newcommand{\msol} {$M_{\odot}$}

\newcommand{\feka} {Fe~K$\alpha$}

\shorttitle{.}
\shortauthors{Brightman et al.}

\begin{document}

\title{A long hard-X-ray look at the dual active galactic nuclei of M51 with {\it NuSTAR}}

\author{M. Brightman$^{1}$, M. Balokovi\'{c}$^{2}$, M. Koss$^{3}$, D. M. Alexander$^{4}$, A. Annuar$^{5}$, H. Earnshaw$^{1}$, P. Gandhi$^{6}$, F. A. Harrison$^{1}$, A. E. Hornschemeier$^{7}$, B. Lehmer$^{8}$, M. C. Powell$^{9}$, A. Ptak$^{10,7}$, B. Rangelov$^{11}$, T. P. Roberts$^{4}$, D. Stern$^{12}$, D. J. Walton$^{13}$, A. Zezas$^{14,15,16}$}

\affil{$^{1}$Cahill Center for Astrophysics, California Institute of Technology, 1216 East California Boulevard, Pasadena, CA 91125, USA\\
$^{2}$Harvard-Smithsonian Center for Astrophysics, 60 Garden Street, Cambridge, MA 02140, USA\\
$^{3}$Eureka Scientific, 2452 Delmer Street Suite 100, Oakland, CA 94602-3017, USA\\
$^{4}$Department of Physics, Centre for Extragalactic Astronomy, Durham University, South Road, Durham DH1 3LE, UK\\
$^{5}$School of Applied Physics, Faculty of Science \& Technology, Universiti Kebangsaan Malaysia, 43600 Bangi, Selangor, Malaysia. \\
$^{6}$School of Physics and Astronomy, University of Southampton, Highfield, Southampton SO17 1BJ, UK\\
$^{7}$NASA Goddard Space Flight Center, Code 662, Greenbelt, MD 20771, USA\\
$^{8}$Department of Physics, University of Arkansas, 226 Physics Building, 825 West Dickson Street, Fayetteville, AR 72701, USA\\
$^{9}$Yale Center for Astronomy and Astrophysics, and Physics Department, Yale University, P.O. Box 2018120, New Haven, CT 06520-8120, USA \\
$^{10}$Johns Hopkins University, Homewood Campus, Baltimore, MD 21218, USA\\
$^{11}$Texas State University, Department of Physics, 601 University Drive, San Marcos, TX 78666, USA\\
$^{12}$Jet Propulsion Laboratory, California Institute of Technology, Pasadena, CA 91109, USA\\
$^{13}$Institute of Astronomy, Madingley Road, Cambridge CB3 0HA, UK\\
$^{14}$Physics Department, University of Crete, 71003 Heraklion, Crete, Greece\\
$^{15}$Harvard-Smithsonian Center for Astrophysics, 60 Garden St., Cambridge, MA 02138, USA\\
$^{16}$Foundation for Research and Technology - Hellas (FORTH), Heraklion 71003, Greece\\}

\begin{abstract}

We present a broadband X-ray spectral analysis of the M51 system, including the dual active galactic nuclei (AGN) and several off-nuclear point sources. Using a deep observation by \nustar, new high-resolution coverage of M51b by \chandra, and the latest X-ray torus models, we measure the intrinsic X-ray luminosities of the AGN in these galaxies.  The AGN of M51a is found to be Compton thick, and both AGN have very low accretion rates ($\lambda_{\rm Edd} <10^{-4}$).  The latter is surprising considering that the galaxies of M51 are in the process of merging, which is generally predicted to enhance nuclear activity.  We find that the covering factor of the obscuring material in M51a is $0.26 \pm 0.03$, consistent with the local AGN obscured fraction at \lx$\sim 10^{40}$ \ergs. The substantial obscuring column does not support theories that the torus, presumed responsible for the obscuration, disappears at these low accretion luminosities. However, the obscuration may have resulted from the gas infall driven by the merger rather than the accretion process.  We report on several extra-nuclear sources with \lx$>10^{39}$ \ergs\ and find that a spectral turnover is present below 10 keV in most such sources, in line with recent results on ultraluminous X-ray sources. 

\end{abstract}


\section{Introduction}
M51, first cataloged by \cite{messier1781}, consists of a pair of interacting galaxies: M51a (NGC 5194), a grand-design spiral galaxy, first to be classified as a spiral galaxy, and M51b (NGC 5195), a dwarf galaxy. The M51 galaxies are among the closest galaxies to our own at a distance of 8.58$\pm$0.10 Mpc, derived from the tip of the red giant branch method \citep{mcquinn16}. As such, the galaxies have become a case study for the effects of galaxy interactions on galaxy evolution. The interaction is believed to be the cause of the distinctive spiral structure of M51a \citep{toomre72}. Several works have investigated the interaction through N-body simulations. While it was originally thought that the two galaxies were experiencing their first close passage, later works have favored multiple past encounters, with one disk-plane crossing 400--500 Myr ago and a more recent one 50--100 Myr ago, both at a separation of $\sim25$ kpc \citep{salo00,theis03,dobbs10}.

While interactions between galaxies have a large effect on their evolution, interactions are also predicted to increase the activity of their central supermassive black holes \citep[SMBHs, e.g.][]{sanders88,hernquist89}. This is due to the massive gas inflows caused by the resulting tidal forces, observational evidence for which has been found in large samples of galaxies \citep[e.g.][]{ellison11,satyapal14,fu18}. The predictions are that the gas inflows into the nuclear regions also obscure the active nucleus \citep{hopkins05}, which has been revealed by recent results \citep{kocevski15,lanzuisi15,ricci17a}. Indeed M51a and M51b are classed as a dual AGN: M51a hosts a Seyfert 2 in its nucleus \citep{stauffer82} that is obscured by Compton-thick (CT) material along the line of sight \citep[\nh$>1.5\times10^{24}$ \cmsq][]{fukazawa01}. Although M51b is classified as a LINER in the optical \citep{devaucouleurs91}, the {\it Spitzer}/IRS detection of [Ne\,{\sc v}]$\lambda14.32 \mu$m confirms that M51b is AGN powered \citep{goulding09}. The AGN in M51b has been estimated to have an X-ray luminosity of $\sim10^{39}$ \ergs\ in the 2--10 keV band \citep{hgarcia14}. There is also evidence for AGN feedback from the nuclei of both galaxies \citep{querejeta16,schlegel16}, a major ingredient in the co-evolution of galaxies and their SMBHs.

The M51 system has been observed by all major X-ray observatories and was first detected in this band by the {\it Einstein Observatory}  \citep{palumbo85}. \chandra\ was the first X-ray observatory to resolve the nucleus of M51a, and found it to have an iron K$\alpha$ line with an equivalent width greater than 2 keV \citep{terashima01,levenson02}. The intrinsic X-ray luminosity has been estimated to be $4\times10^{40}$ \ergs\ \citep{xu16} in the 2--10 keV band, which makes it the lowest luminosity CTAGN known. 

Besides detecting both nuclei, the {\it Einstein Observatory} found several ultraluminous X-ray sources \citep[ULXs, see][for reviews]{roberts07,kaaret17} associated with the galaxies. The ULX population was resolved into eight different sources by the high-resolution imager on {\it ROSAT} \citep{ehle95}. The ULXs have since been extensively cataloged and characterized by \cite{liu05}, \cite{dewangan05}, \cite{winter06}, \cite{swartz11}, \cite{walton11} and \cite{kuntz16}. \cite{kuntz16} found that typically $\sim$5 ULXs were active at any one time, with only two persistently active over 12 years of \chandra\ observations. The large number of ULXs in M51 is likely related to the high rates of star formation \citep[$\sim$2.6\msol\,yr$^{-1}$][]{schuster07} that were triggered by the interaction between the galaxies \citep{smith12}. The ULXs in M51 show several interesting properties such as two eclipsing ULXs \citep{urquhart16}, an intermediate-mass black hole (IMBH) candidate \citep{earnshaw16}, one that demonstrates apparent bimodal flux behaviour that could indicate a pulsar in the propeller regime \citep{earnshaw18} and one where a cyclotron resonance scattering feature has been detected, identifying the source as a neutron-star accretor \citep{brightman18}.

In this paper we present a 210 ks observation of M51 with the {\it Nuclear Spectroscopic Telescope Array} \citep[\nustar,][]{harrison13} with the aim of characterizing the spectra of the nuclei of the two galaxies and the extranuclear point source population above 10 keV. While a short 18 ks \nustar\ observation of M51 was already analyzed and presented by \cite{xu16}, who studied the nucleus of M51a, and \cite{earnshaw16}, who investigated one of the ULXs, the observation was too short to constrain spectral parameters well. In addition to the \nustar\ data, we present a 37.8 ks \chandra\ ACIS-I observation that was taken contemporaneously and provides both soft X-ray coverage and the angular resolution to separate the crowded field. Furthermore, many \chandra\ ACIS-S observations already exist on M51 \citep{kuntz16,lehmer17}, these observations have all placed the nucleus of M51b at $\sim4$\arcmin\ off-axis where the PSF is degraded and the source is near the edge of the detector or off it completely. Here we use ACIS-I with the aim point between the galaxies in order to better resolve the nucleus of M51b.

\section{Observational data reduction and analysis}
\label{sec_obs}

The 37.8-ks \chandra\ and 210-ks \nustar\ observations studied here were taken contemporaneously between 2017 March 16$-$17. Table \ref{table_obsdat} provides a description of the observational data. In addition to these data, we use archival \chandra\ data taken during a large program with this observatory in 2012 in order to inform us of the long term flux behavior of these sources \citep{kuntz16}. The details of these observations are listed in Table \ref{table_obsdat}. The following sections describe the individual observations and data reduction. Spectral fitting was carried out using {\sc xspec} v12.9.1 \citep{arnaud96} and all uncertainties quoted are at the 90\% level.

\begin{table}
\centering
\caption{Observational data}
\label{table_obsdat}
\begin{center}
\begin{tabular}{l l c r}
\hline
Observatory	& ObsID	& Start date (UT)	& Exposure \\
			&		&				& (ks)	   \\
\hline
\chandra\	& 13813			& 2012-09-09 17:47:30	& 179.2 \\
\chandra\	& 13812			& 2012-09-12 18:23:50	& 179.2 \\
\chandra\	& 15496			& 2012-09-19 09:20:34	& 179.2 \\
\chandra\	& 13814			& 2012-09-20 07:21:42	& 179.2 \\
\chandra\	& 13815			& 2012-09-23 08:12:08	& 179.2 \\
\chandra\	& 13816			& 2012-09-26 05:11:40	& 179.2 \\
\chandra\	& 15553			& 2012-10-10 00:43:36	& 179.2 \\
\nustar\	& 60201062002	& 2017-03-16 15:21:09	& 47.2  \\
\chandra\	& 19522			& 2017-03-17 00:48:01	& 37.8 \\
\nustar\	& 60201062003	& 2017-03-17 16:56:09	& 163.1 \\
\hline
\end{tabular}
\end{center}
\end{table}

\subsection{Chandra}
\label{subsec_chandra}

The primary \chandra\ observation of M51 used here, obsID 19522, was taken with ACIS-I at the optical axis. The aim point was placed between the two galaxies in order to improve upon the PSF size at the location of M51b, which in prior observations has largely been placed at large off-axis angles. We use these \chandra\ data to resolve the sources in the galaxy at $\sim$arcsec scales and to provide contemporaneous soft X-ray coverage to the \nustar\ data. The \chandra\ data were also used to inform the positions of the \nustar\ sources and were analyzed with {\sc ciao} v4.7. 

No formal method is employed here to select the sources for which we conduct joint spectral fitting with \nustar. However, our informal method is to select sources which are bright in the 3--8 keV band as indicated by \chandra, that also show indications for emission in the \nustar\ 3--8 keV image. To do this, we first use the {\sc ciao} tool {\sc dmcopy} to create a 3--8 keV \chandra\ image which we show in Figure \ref{fig_multiband_img}. For clarity, we plot these sources as contours which represent sources with $>2$ counts pixel$^{-1}$. Also plotted in Figure \ref{fig_multiband_img} with the same scale is the {\it Spitzer}/IRAC 3.6\micron\ image for comparison. 

\subsection{NuSTAR}
\label{subsec_nustar}

\begin{figure*}
\begin{center}
\includegraphics[width=180mm]{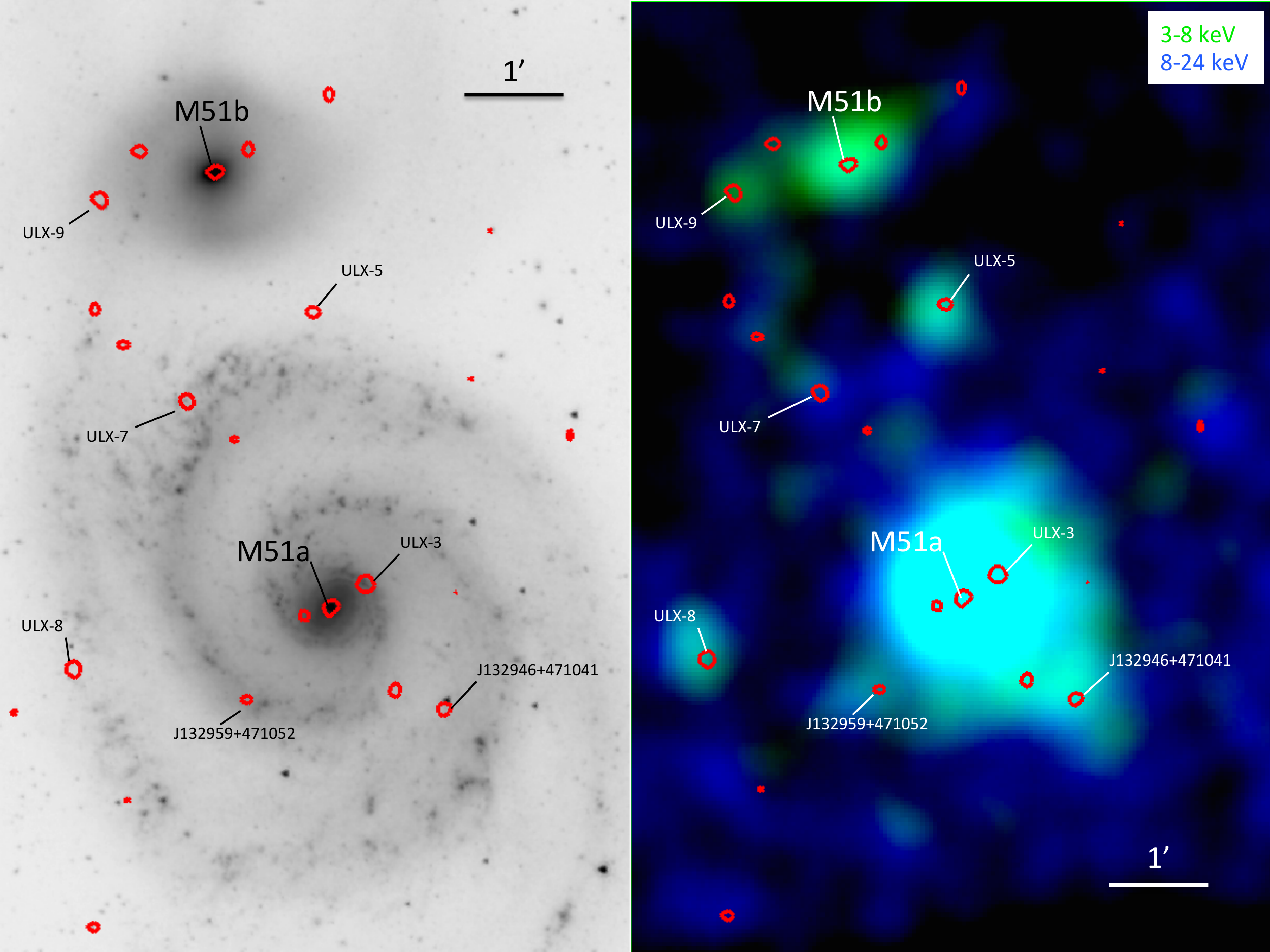}
\caption{Multiband images of M51. The panel on the left shows {\it Spitzer}/IRAC 3.6 \mic\ image, overlaid with \chandra\ 3-8 keV contours, showing the brightest sources ($>2$ counts pixel$^{-1}$), in red. The right panel shows the same contours overlaid on the  \nustar\ 3--8 keV image in green and the \nustar\ 8--24 keV image in blue. The \nustar\ images, which consist of co-added FPMA+B images, have been smoothed with a 10\arcsec\ kernel. }
\label{fig_multiband_img}
\end{center}
\end{figure*}

The raw \nustar\ data consist of two obsIDs, 60201062002 and 60201062003, which have slightly different pointings, and were reduced using the {\sc nustardas} software package version 1.7.0. The events were cleaned and filtered using the {\tt nupipeline} script with standard parameters. Due to higher-than-usual background during passages of the South Atlantic Anomaly (SAA), we filter the events files using {\tt saacalc=1} and {\tt saamode=strict}. We also use {\sc xselect} to make images in the 3--8 keV and the 8--24 keV bands and use {\sc ximage} to co-add the images from the two obsIDs and the two focal plane modules (FPMs). Based on the positions of the brightest sources, we found an astrometric offset of 8.5\arcsec\ between the \chandra\ and \nustar\ positions. The offset was $\Delta$RA=7.4\arcsec\ and $\Delta$Dec=4.3\arcsec, which are typical values for astrometric offsets between \chandra\ and \nustar\ \citep{lansbury17}. We show the astrometrically corrected images in Figure \ref{fig_multiband_img}.

Figure \ref{fig_multiband_img} shows that \nustar\ finds $\sim8$ sources of emission within the galaxies of M51 in the 3--8 keV band, for some of which the PSFs are overlapping. Furthermore, the high resolution \chandra\ data show that for at least two \nustar\ sources, more than one source contributes significantly to the \nustar\ PSF. We select these eight sources for spectral analysis and conduct joint spectral fitting of the sources where \chandra\ resolved more than one source. Approximately, these sources are selected with 3--8 keV fluxes $>10^{-14}$ \ergcms. From Poisson statistics, we find that all sources are significantly detected in the 0.5--8 keV \chandra\ and 3--30 keV \nustar\ bands, with the probabilities of background fluctuations being $<<10^{-10}$.

We use the {\tt nuproducts} task to generate the spectra and the corresponding response files for each obsID separately. For the brightest \nustar\ source, the nucleus of M51a, we extract spectra using a 40\arcsec\ circular region,  which corresponds to an encircled energy fraction of $\sim55$\% \citep{harrison13,madsen15}. This extraction region includes ULX3, so we fit the spectra of these two sources jointly. For the rest of the \nustar\ sources, which are fainter, we use 20\arcsec\ circular regions in order to reduce background and avoid overlapping PSFs. 20\arcsec\ encloses around 30\% of the counts. Despite this, the extraction region for the nucleus of M51b includes at least one other source that contributes significantly to the \nustar\ PSF, so we model the spectra of these sources jointly. Background spectra were extracted from 50\arcsec\ circular regions on the same detector as the respective source taking care to avoid detected sources, including those not evident in the \nustar\ image but shown to be bright in the 3--8 keV band by \chandra. For each source, the spectral data from FPMA for each observation were co-added to each other using the {\sc heasoft} tool {\sc addspec}. The same was done for the data from FPMB. The data from FPMA and FPMB were not co-added to each other, but used for simultaneous fitting instead. The resulting exposure after background filtering and co-adding is 193.8 and 193.3 ks for FPMA and FPMB respectively.

\subsection{Palomar}

Optical spectroscopic observations of the nucleus of M51a were taken as part of the BAT AGN Spectroscopic Survey (BASS) followup of newly detected 105-month \swiftbat-detected AGN \citep{oh18}. The aim was to determine the mass of the SMBH from the stellar velocity dispersion. The observation took place at UT 2018 March 27 with the Palomar Double Spectrograph (DBSP) on the Hale 200-inch Hale telescope for 600 s using the 1200 lines/mm grating. We used the 2\arcsec\ slit at the parallactic angle (160\degree) and extracted a nuclear aperture of 10\arcsec. We measured the sky lines to have a FWHM=2.4\,\AA\ at 5000\,\AA\ and FWHM=2.1\,\AA\ at 8500\,\AA\ corresponding to instrumental limit of 64 \kms\ and 30 \kms\ respectively, at the redshift of M51.

The velocity dispersion was measured using the penalized PiXel Fitting software \citep[{\tt pPXF};][]{cappellari04, cappellari17} version 5.0 to fit with the optimal stellar templates. We used 86 stars from The X-Shooter Spectral Library of stellar spectra \citep{chen14} with R=10,000 which cover 3000-25000 \AA. These templates have been observed at higher spectral resolution than the AGN observations and are convolved in {\tt pPXF} to the spectral resolution of each observation before fitting. When fitting the stellar templates all of the prominent emission lines were masked. After the best fitting stellar template was removed, the residual emission lines were fit. We found a velocity dispersion of 75$\pm$4 \kms\ when fitting the 3950\AA\ to 5500\AA\ which includes the CaH+K and Mg\,{\sc i} regions and 63$\pm$4 \kms\ when fitting the calcium triplet region (8350-8900\AA). More details on the reductions and pPXF fitting can be found in \cite{koss17}. We note both these measurements are significantly below the past literature value \citep[102 \kms,][]{nelson95}, most likely due to the uncertainties in subtracting the instrumental resolution of lower resolution observations.

\section{Spectral fitting}
\label{sec_specfit}
We carry out X-ray spectral fitting on the eight brightest \nustar\ sources in M51: the nucleus of M51a, which includes the emission from ULX3; the nucleus of M51b, which includes emission from a bright extra-nuclear source; and six additional off-nuclear sources. We describe the spectral fitting procedures for each source individually in the following subsections. The count rate for each source from each detector are shown in Table \ref{table_cntrts}. 

For the bright nucleus of M51a, we group both the \chandra\ and \nustar\ data with a minimum of 20 counts. We carry out spectral fitting with background subtracted spectra and use the \chisq\ statistic as the fit statistic. The \nustar\ data remain source dominated up to $\sim30$ keV, but we consider the whole 3--79 keV \nustar\ band for spectral fitting. We carry out spectral fitting over the 0.5--8 keV band for the \chandra\ data. 

\begin{figure}
\begin{center}
\includegraphics[width=90mm]{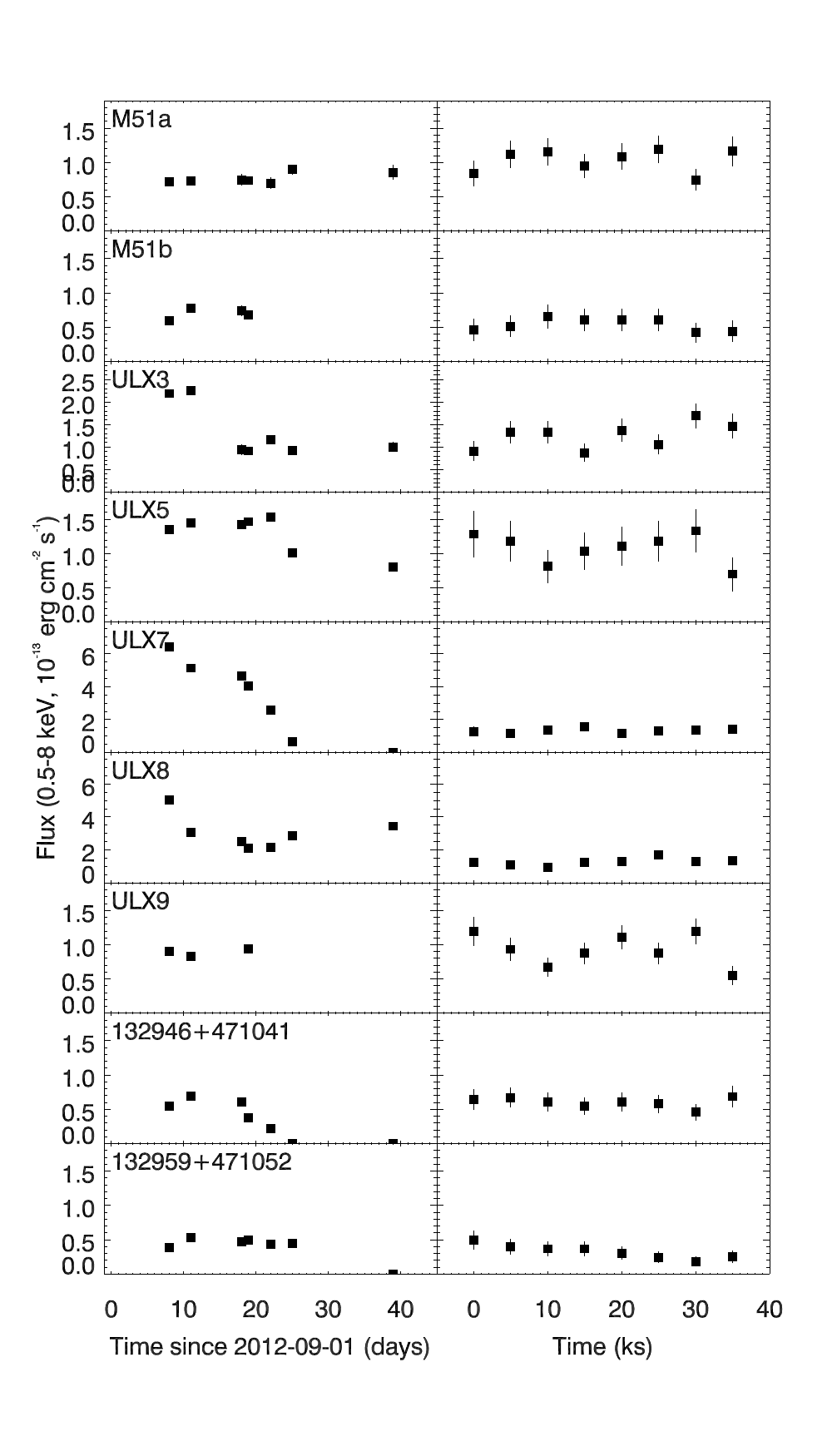}
\caption{Long-term (left) and short-term (right) \chandra\ lightcurves of the sources studied here. Fluxes in the 0.5--8 keV band are converted from count rates assuming the best-fit spectral models.}
\label{fig_M51_ltcrv}
\end{center}
\end{figure}

Since many archival \chandra\ observations of M51 are available in addition to the one simultaneously taken with \nustar\, we investigate the long-term flux behavior of our sources. Specifically we study a period in 2012 where many of these observations were taken over a period of $\sim40$ days. We also examine the variability of each source during the latest \chandra\ observation. For this, we convert the background subtracted count rates into 0.5--8 keV fluxes assuming the spectral models described in the following sections. We plot the long and short-term lightcurves in Figure \ref{fig_M51_ltcrv}.

The long-term 0.5--8 keV flux of the M51a nucleus remains approximately constant, varying by $\pm20$\% around the mean.  Since it has not varied considerably over the years, we utilize all existing \chandra\ data on the nucleus of M51a. These are obsIDs 13812, 13813, 13814 , 13815, 13816, 15496, 15553 and 19522, totaling 783 ks of exposure.  We fit the \chandra\ data and the \nustar\ data simultaneously using cross calibration constants, $C_{\rm ACIS}$, $C_{\rm FPMA}$, and $C_{\rm FPMB}$  to account for the differing instrumental responses. $C_{\rm FPMA}$ is fixed to unity while the others are free to vary.

For the nucleus of M51b, the differing off-axis angles and PSF sizes that have resulted from these in the previous observations make assessing the variability challenging so we do not co-add the \chandra\ data of M51b. Most of the ULXs show long term flux variability. Therefore for these sources we only use the latest \chandra\ obsID 19522 that was simultaneous with \nustar. Also, due to their low-count nature, we only lightly group the spectra, with a minimum of 1 count for \chandra\ and 3 counts for \nustar\ \citep[see][for an investigation into the grouping of spectra with a low number of counts per bin]{lanzuisi13}. We use the Cash statistic \citep{cash79} with background subtracted spectra. While the use of the Cash statistic cannot be strictly used in the case where the background has been subtracted, {\sc xspec} implements a modified version of the Cash statistic to account for this, known as the W-statistic\footnote{https://heasarc.gsfc.nasa.gov/xanadu/xspec/manual/XSappendixStatistics.html}. We only consider \nustar\ data up to 30 keV, beyond which the photon statistics are poor.  For these sources we fix all cross calibration constants to unity unless we find evidence for a significant deviation from this. 

\begin{table}
\centering
\caption{X-ray count rates for all the sources in M51 analyzed here}
\label{table_cntrts}
\begin{center}
\begin{tabular}{l c c c}
\hline
Source	& ACIS 			& FPMA 					& FPMB \\	
		& 0.5--8 keV		& 3--30 keV 				& 3--30 keV \\
\hline
M51a nucleus	& 3.5$\pm$0.1	& \multirow{2}{*}{7.7$\pm$0.2} & \multirow{2}{*}{7.5$\pm$0.2} \\
ULX3		& 3.7$\pm$0.3\\
M51b nucleus	& 2.2$\pm$0.2	& \multirow{2}{*}{0.5$\pm$0.1} & \multirow{2}{*}{0.7$\pm$0.1} \\
extra-nuclear source	& 0.8$\pm$0.2\\
ULX5		& 2.6$\pm$0.3	& 0.6$\pm$0.1				& 0.5$\pm$0.1 \\
ULX7		& 4.4$\pm$0.3	& 0.2$\pm$0.1				& 0.3$\pm$0.1 \\
ULX8		& 8.3$\pm$0.5	& 0.5$\pm$0.1				& 0.5$\pm$0.1 \\
ULX9		& 5.7$\pm$0.4	& 0.4$\pm$0.1				& 0.3$\pm$0.1 \\
J132946+471041& 3.6$\pm$0.3	& 0.7$\pm$0.1				& 0.4$\pm$0.1 \\
J132959+471052& 1.9$\pm$0.2	& 0.4$\pm$0.1				& 0.3$\pm$0.1 \\

\hline
\end{tabular}
\tablecomments{Background-subtracted source count rates (counts ks$^{-1}$) in the \chandra\ ACIS-I, \nustar\ FPMA, and \nustar\ FPMB detectors.}

\end{center}
\end{table}

\subsection{The Compton-thick nucleus of M51a}
\label{subsec_m51a}

The nucleus of M51a is well known to be Compton thick, that is, the optical depth to Compton scattering of X-rays off electrons is greater than unity. Therefore the effect of Compton scattering must be taken into account when modeling the X-ray spectra of these sources. Several X-ray spectral models now exist that have been compiled especially for this reason. These models take an intrinsic AGN power-law spectrum and subject it to photo-electric absorption, Compton scattering and iron fluorescence using Monte-Carlo simulations, assuming a toroidal obscuring structure thought to exist in the inner regions of AGN.

We use the most up to date X-ray spectral torus models in our analysis, the {\tt mytorus} model \citep{murphy09} and the {\tt borus}\footnote{The specific geometry and version number used is {\tt borus02\_afe1\_v170227a.fits}} model \citep{balokovic18}, which is an update to the {\tt torus} model \citep{brightman11}. These models differ mostly in the geometry of the toroidal structure assumed. Both assume a smooth axisymmetric structure, but for {\tt mytorus}, the torus has a circular cross section, whereas the {\tt borus} model is based on a sphere with a biconical cut out. In this way, for lines of sight through the torus, the line-of-sight-\nh\ in the {\tt mytorus} is inclination dependent, but not for the {\tt borus} model.  For both models, the direct transmitted component, which is composed of photons that travel through the structure without undergoing interactions, can be decoupled from the scattered component, where the photons have Compton scattered off the obscuring material. This gives the models greater freedom, such as to emulate different geometries, like a clumpy distribution of gas, but also adds degeneracies. For X-ray spectral fitting of absorbed sources, there is a degeneracy between the X-ray spectral index, $\Gamma$, and the column density, \nh, especially in low quality spectra over a narrow band. This degeneracy is mostly mitigated using these torus models that include the fluorescent lines and when high quality broad-band spectra are used, such as we have here. For the models with the covering factor as a free parameter, this can be degenerate with the \nh, but again this mostly affects lower signal to noise data \citep[e.g.][]{balokovic18}.

\begin{figure}
\begin{center}
\includegraphics[width=90mm]{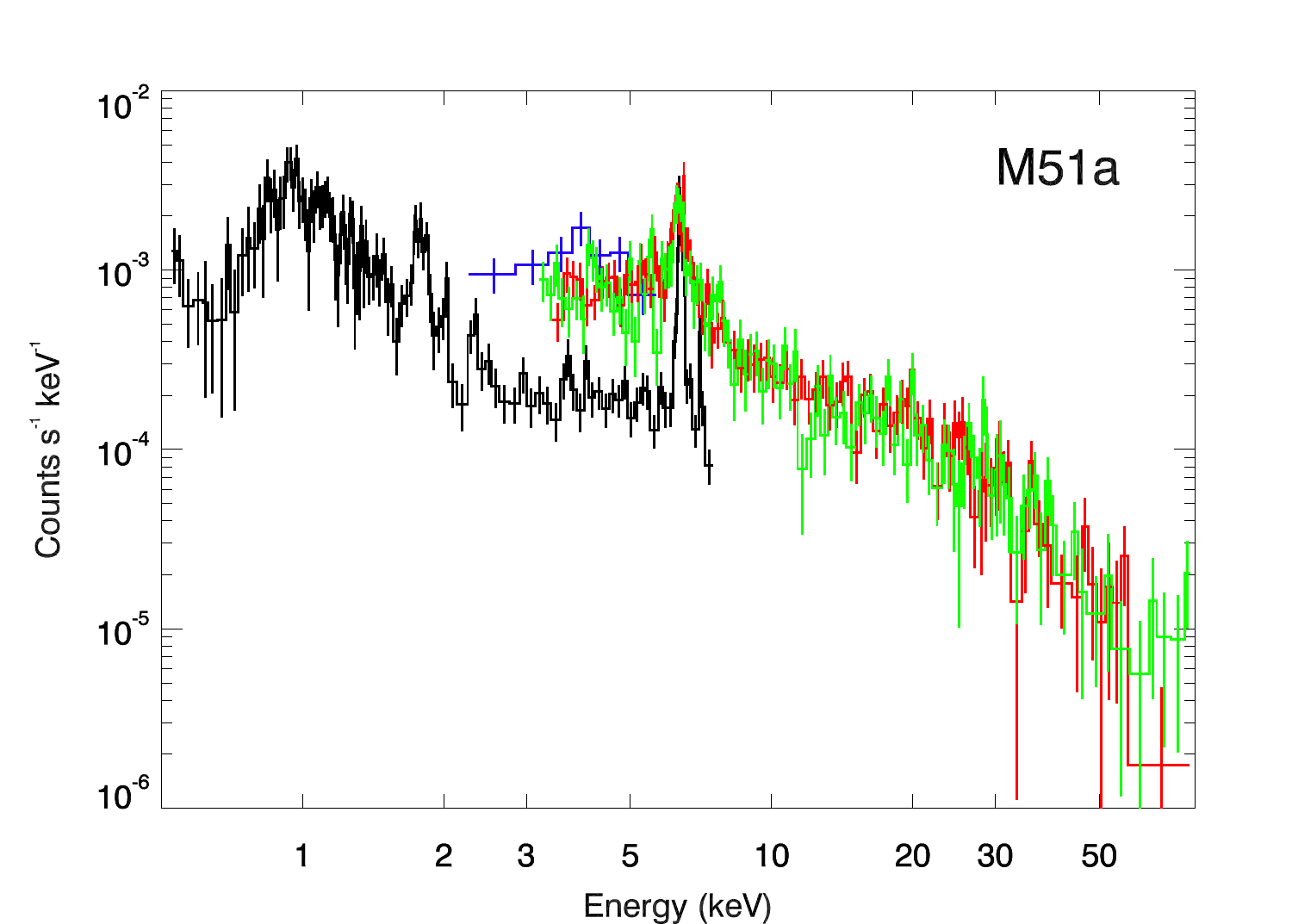}
\caption{\chandra\ (black is the nucleus and blue is ULX3) and \nustar\ (FPMA in red and FPMB in green) spectra of the M51a nuclear region folded through the instrumental responses, and rebinned for clarity.}
\label{fig_M51a_spec}
\end{center}
\end{figure}

For spectral fitting of the nucleus of M51a we use the co-added \chandra\ spectra extracted and merged with {\sc acis extract} \citep{broos10}. We only consider emission from within 1\arcsec\ of the central point source, corresponding to an encircled energy fraction of 90\%, and ignore the extended emission surrounding the nucleus since this does not contribute in the \nustar\ band above 3 keV as seen in the 3--8 keV \chandra\ image (Figure \ref{fig_multiband_img}). See \cite{xu16} for a description of the soft extended emission. The 40\arcsec\ region used to extract the \nustar\ data also includes two hard point sources seen in the \chandra\ images. One of these is ULX3, the \chandra\ spectrum of which we include in our spectral analysis. We model the spectrum of ULX3 with an absorbed cut-off power-law. The other hard source is not bright enough to contribute significantly to the \nustar\ spectrum, with less than 1/3 the count rate of the nucleus or ULX3 in the 3--8 keV band, so we neglect it in spectral fitting. The spectral data are shown in Figure \ref{fig_M51a_spec}.

Despite only extracting spectra from within the central 1\arcsec, which corresponds to $\sim40$ parsecs at 8.5 Mpc, excess soft X-ray emission is seen in the \chandra\ spectrum above the characteristic reflection spectrum. This likely arises from scattered nuclear light and emission from gas photoionized by the AGN. In order to avoid the complexities associated with this emission, we consider only counts above 3 keV.

We find that {\tt mytorus} in coupled mode (\nh, $\Gamma$, inclination and normalization of the transmitted component and scattered component tied to each other) provides a fit to the data with \chisq=399.6 with 258 degrees of freedom (DoF). In order to find a possible better fit, we decouple the inclination of the transmitted component and the scattered component, which improves the fit statistic to \chisq/DoF=374/257. We then decouple the \nh\ parameter of the two components which leads to an improvement of the fit statistic to \chisq/DoF=356.9/256. Finally we decouple the normalizations, which leaves the two components fully decoupled, however, this actually worsens the fit (\chisq/DoF=363.0/255). We therefore declare our best-fit {\tt mytorus} model to be the one where only the inclination and \nh\ parameters of the transmitted and scattered components are decoupled, with \chisq/DoF=356.9/256.

For the {\tt borus} model, we start by constraining the inclination of the scattered component to be greater than the opening angle of the torus (such that the line of sight is through the torus), to be fully self consistent. This provides a fit to the data with \chisq=359.3 with 258 DoF. However, if we allow the full range of inclination angles for the scattered component, we find the data prefer a small, close to face-on view of the torus for the scattered component, improving the fit significantly to \chisq/DoF=315.7/258. This was similarly the case for the {\tt mytorus} model, and implies that the observed scattered component is stronger than the simple geometries of the torus models can account for.  Decoupling the \nh\ parameter of the {\tt borus} model does not improve the fit (\chisq/DoF=318.4/256). Likewise is true when decoupling the normalization parameter (\chisq/DoF=318.4/255). We therefore declare our best-fit {\tt borus} model to be the one where only the full range of inclination angles for the scattered component is allowed, with \chisq/DoF=315.7/258.

\begin{figure}
\begin{center}
\includegraphics[width=90mm]{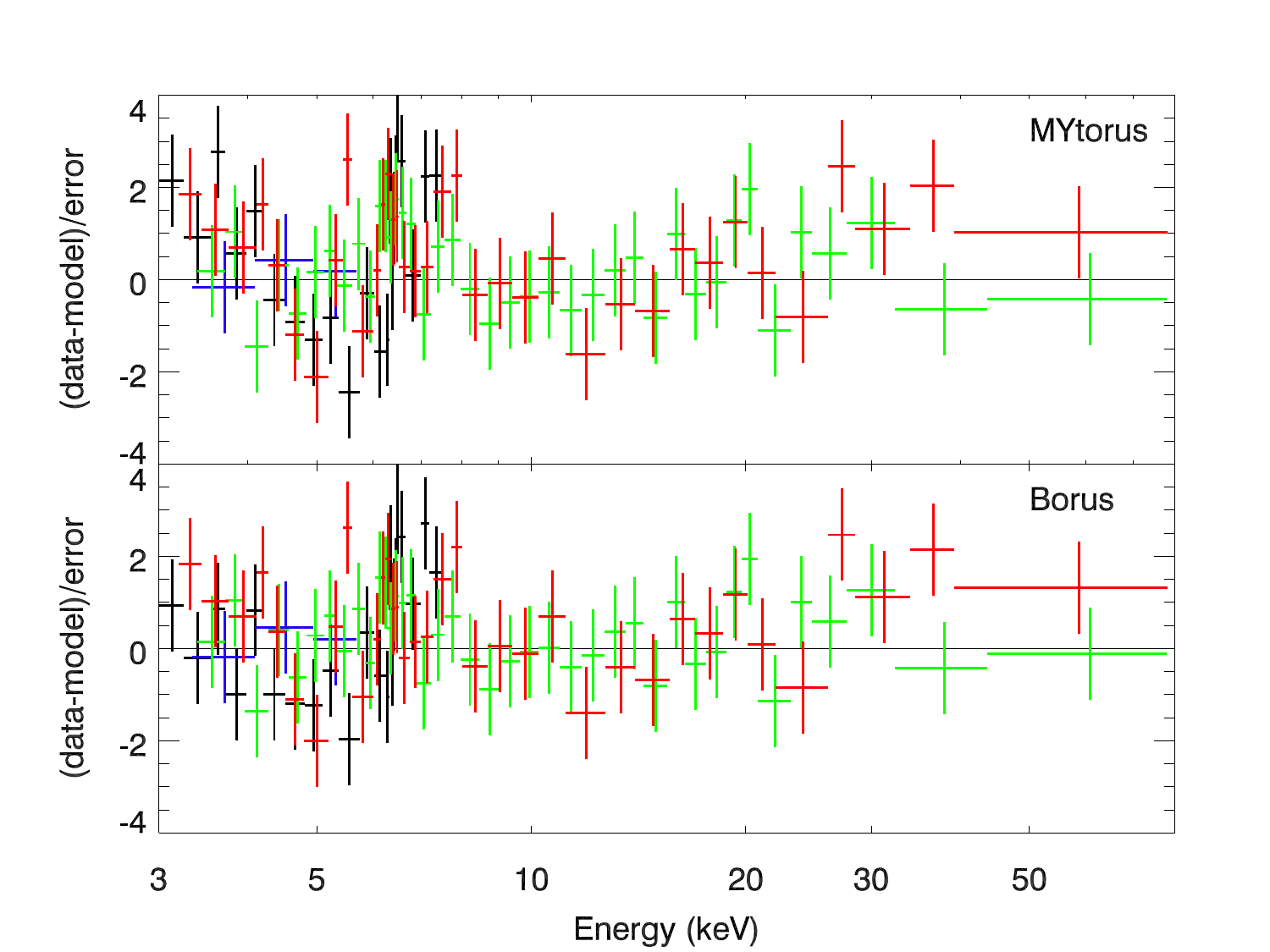}
\caption{\chandra\ (black is the nucleus and blue is ULX3) and \nustar\ (FPMA in red and FPMB in green) residuals to the fits with the {\tt mytorus} and {\tt borus} models. Included in both fits is the {\tt cutoffpl} model for ULX3.}
\label{fig_M51a_spec2}
\end{center}
\end{figure}

\begin{figure}
\begin{center}
\includegraphics[width=90mm]{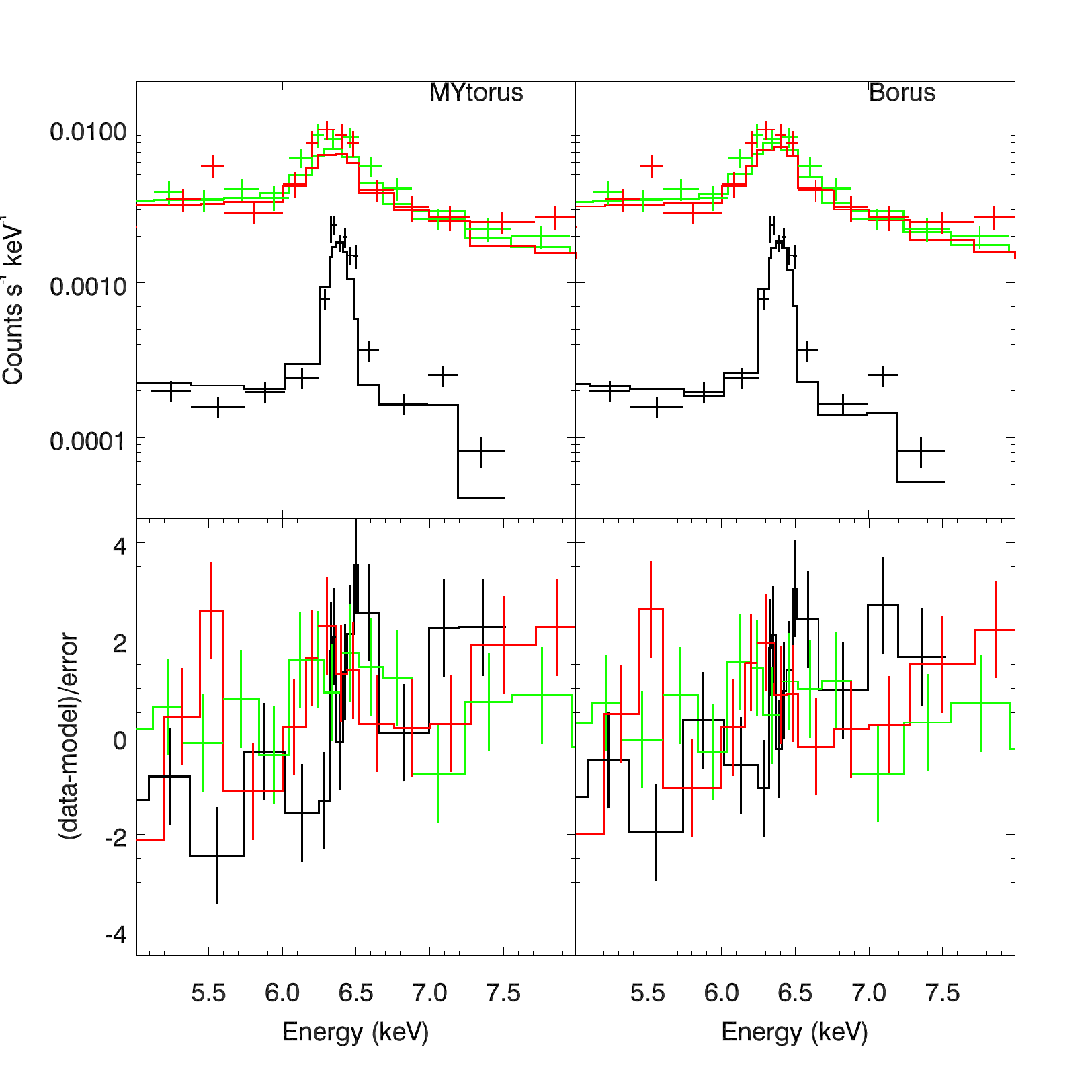}
\caption{Top - \chandra\ (black) and \nustar\ (FPMA in red and FPMB in green) spectra of the M51a nucleus around the \feka\ emission line. The \nustar\ data have been multiplied by a factor of 4  for plotting clarity. Bottom - data to model ratios.}
\label{fig_M51a_spec3}
\end{center}
\end{figure}

We show the broadband residuals to the fit with both the best-fit {\tt mytorus} and {\tt borus} models in Figure \ref{fig_M51a_spec2}. These residuals do not show any obvious deviations with the exception of the Fe-K complex around 6.4 keV. We show a zoom in of the 5--8 keV band which contains the Fe-K complex in Figure \ref{fig_M51a_spec3}. Considering the spectral data in this energy band alone, \chisq/DoF=156.6/64 for {\tt mytorus} and \chisq/DoF=133.0/66 for {\tt borus}, which shows that neither model fits the \feka\ line well, although {\tt borus} provides a slightly better fit.

The equivalent width (EW) of the \feka\ line of M51a is known to be one of the highest measured \citep{terashima01,levenson02}. The torus models that we utilize here treat the \feka\ and continuum self-consistently and as such the \feka\ line plays a part in constraining the torus geometry. Since the \feka\ line is not treated as a separate component in our fits a direct measurement of the EW is not given from these models. However, for {\tt mytorus} the lines component can be decoupled from the continuum. We measure the EW of the \feka\ line by decoupling the normalization of the lines component in {\tt mytorus}, which yields 4.1 keV. This is for the \feka\ line, however, this still includes constraints from the Fe K$\beta$ line in the fit, which cannot be remove from the spectral fit. We then remove the fluorescent lines altogether from the {\tt mytorus} fit and add a single Gaussian component at 6.4 keV to model the \feka\ emission. For this Gaussian component, we measure E$=6.41\pm0.01$ keV, $\sigma=51^{+12}_{-13}$ eV and EW$=3.3^{+0.27}_{-0.46}$ keV. The lower EW measured by the Gaussian with respect to {\tt mytorus} may be due to the Fe K$\beta$ line being stronger with respect to the \feka\ line than the model expectation.

For the {\tt borus} model of the nucleus, $C_{\rm ACIS}/C_{\rm FPMA}=0.84^{+0.12}_{-0.06}$ and for the {\tt cutoffpl} model of the ULX $C_{\rm ACIS}/C_{\rm FPMA}=0.69^{+0.12}_{-0.11}$. While the cross-calibration constant for the nucleus is almost consistent with unity, the one for the ULX is not. The \chandra\ lightcurve of the ULX shows that it is rising in flux (Figure \ref{fig_M51_ltcrv}), which it may have continued to do during the latter part of the \nustar\ observation when \chandra\ was no longer observing. This would explain the apparent greater contribution of the ULX to the \nustar\ spectrum than \chandra\ shows. Similar cross-calibration constants result from the {\tt mytorus} model.

\begin{figure}
\begin{center}
\includegraphics[width=90mm]{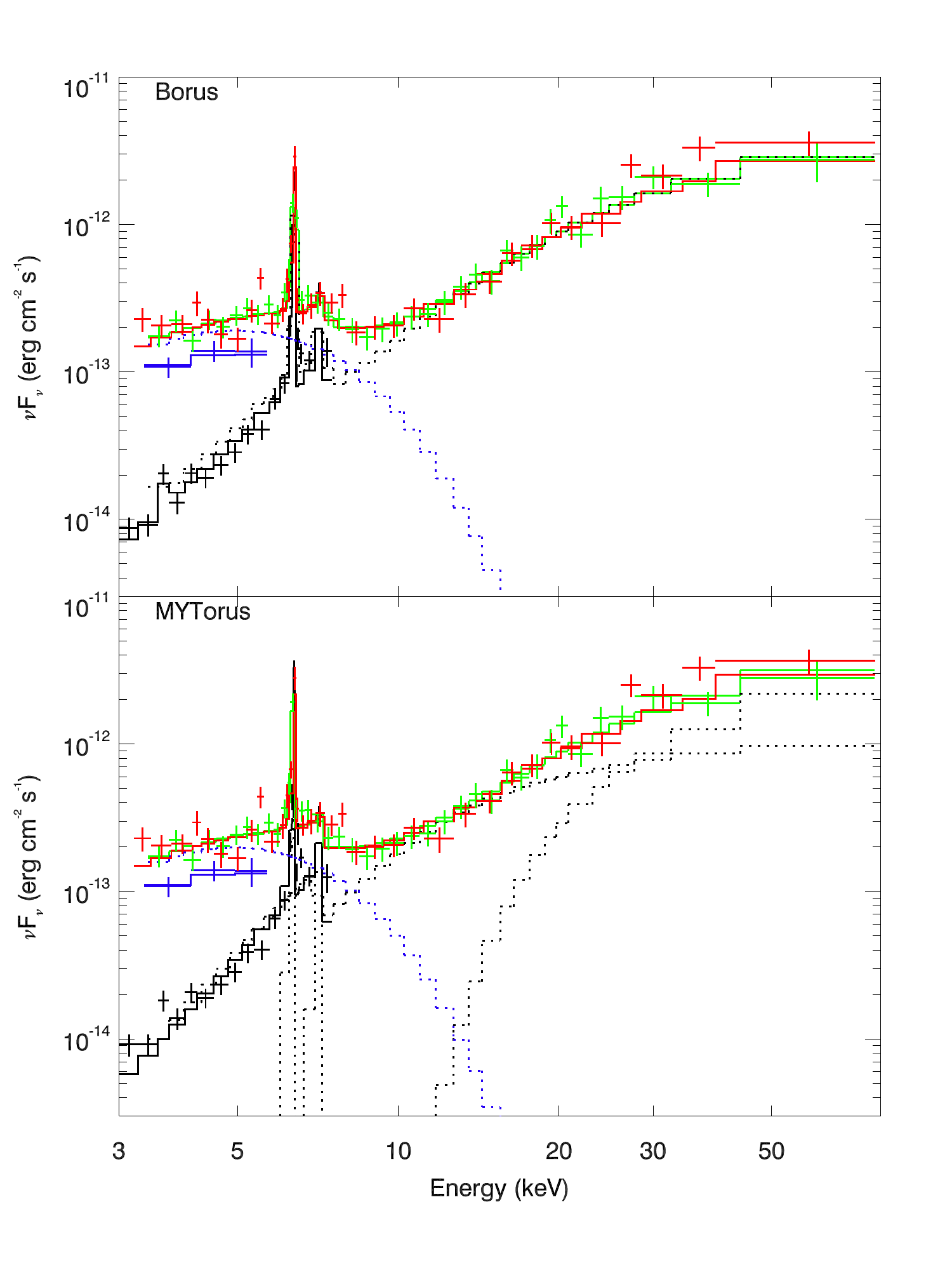}
\caption{\chandra\ (black is the nucleus and blue is ULX3) and \nustar\ (FPMA in red and FPMB in green) spectra of the M51a nuclear region unfolded through the instrumental responses assuming the best-fit {\tt borus} model (top, black dotted line) and {\tt mytorus} model (bottom, black dotted line showing transmitted, scattered and lines components) for the nucleus and {\tt cutoffpl} model for ULX3 (blue dotted line).}
\label{fig_M51a_spec4}
\end{center}
\end{figure}

We show the best-fit {\tt borus} model of the nucleus, and {\tt cutoffpl} model of ULX3, in Figure \ref{fig_M51a_spec4}. We list the best-fit parameters for both the best-fit {\tt mytorus} and {\tt borus} model fits to the nucleus in Table \ref{tab_m51a_specpar}. We present best-fit parameters for ULX3 in Table \ref{tab_specpar} along with the other extra-nuclear sources. The \nh\ of the nucleus measured by the two torus models agree, showing that the source is heavily Compton-thick (although {\tt mytorus} does not probe log(\nh/\cmsq)$>25$). Both models also indicate that the intrinsic photon index is low, $\Gamma=1.4-1.8$, (the hard lower limit on both models is 1.4). The AGN is estimated to have an intrinsic 2--10 keV X-ray luminosity of 2$\times10^{40}$ \ergs\ from both models. The {\tt borus} model also estimates a low covering factor of the torus of 0.26$\pm0.03$.

\begin{table*}
\centering
\caption{M51a X-ray spectral parameters}
\label{tab_m51a_specpar}
\begin{center}
\begin{tabular}{l l l l l l l}
\hline
\nh & $\Gamma$ & $f_{\rm C}$ & $\theta_{\rm i}$ & \chisq/DoF & \fx\ & \lx \\
(1) & (2) & (3) & (4) & (5) & (6) & (7)  \\
\hline
\multicolumn{7}{l}{\bf \tt MYTorus}\\

24.9$^{+u}_{-0.2}$ ,  24.0$^{+0.3}_{- 0.2}$ & 1.58$^{+ 0.20}_{-l}$ & 0.5 &90.00$^{+ u}_{-l}$ ,  4.6$^{+u}_{-l}$ &356.9/256 &  2.0$^{+  1.1}_{-  0.7}$ &  1.8$^{+  1.0}_{-  0.7}$ \\
\multicolumn{7}{l}{\bf \tt borus}\\
25.3$^{+u}_{- 0.4}$ & $<$1.43 & 0.26$\pm0.03$ & $<$22 &315.7/258&  2.0$^{+  0.1}_{-  0.6}$ &  1.7$^{+  0.1}_{-  0.5}$ \\

\hline
\end{tabular}
\tablecomments{The best-fit parameters for the {\tt mytorus} and {\tt borus} models to the \chandra\ and \nustar\ spectrum of the nucleus of M51a. Where `$+u$' is indicated, the parameters has hit the upper bound in error estimation. Likewise `$-l$' indicates it has hit its lower bound. Column (1) gives the logarithm of the column density, \nh\ in units of \cmsq. For {\tt mytorus} model, the \nh\ of the transmitted and the scattered components are both listed. Column (2) shows the intrinsic power-law index for each model. Column (3) gives the covering factor of the torus for each model. For {\tt mytorus} this is not a free parameter and fixed at 0.5. Column (4) lists the inclination derived from each model, again, the transmitted and scattered components are both given for {\tt mytorus}. Column (5) lists the \chisq\ for each fit and the number of degrees of freedom. Column (6) gives the intrinsic 2--10 keV flux in units of $10^{-12}$ \ergcms, and Column (7) lists the intrinsic luminosity (corrected for absorption) in units of $10^{40}$ \ergs\ in that band assuming a distance of 8.58 Mpc.}

\end{center}
\end{table*}

\subsection{The LINER nucleus of M51b}
\label{subsec_m51b}

\begin{figure*}
\begin{center}
\includegraphics[width=180mm]{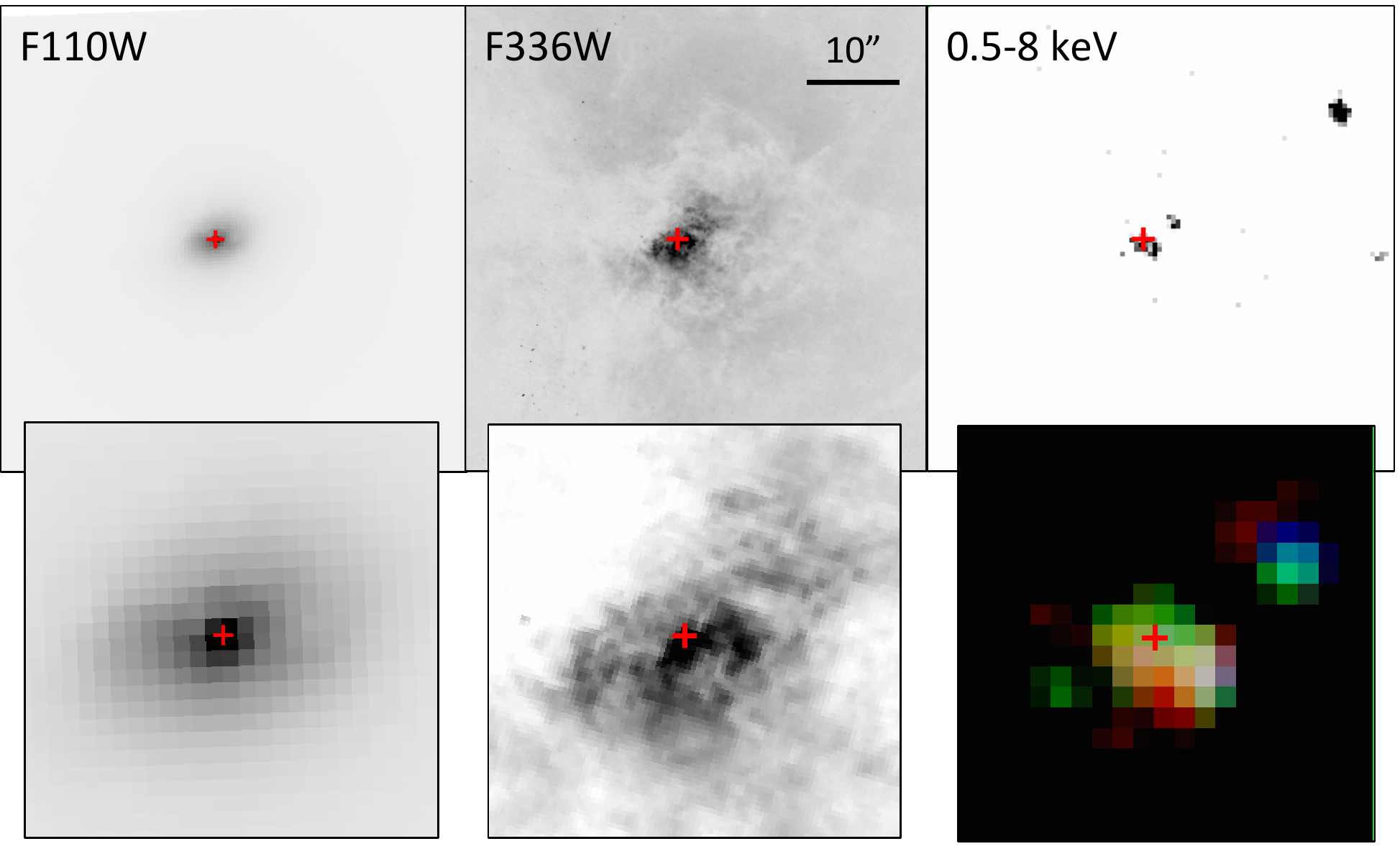}
\caption{Multiband images of the nuclear region of M51b. From left to right these show the  {\it HST}/WFC3 F110W image (peak wavelength 1.15 microns), the {\it HST}/WFC3 F336W (peak wavelength 337.5 nm), and the \chandra\ 0.5--8 keV image. The red cross marks the NIR position of the nucleus. Images in the top panels are 50\arcsec\ on a side, while the images on the bottom rows are a zoom in to the central 10\arcsec. Here we show the \chandra\ image in three bands, 0.5--1.2 keV (red), 1.2--2.5 keV (green) and 2.5--8 keV (blue), which have been smoothed with a Gaussian of 1\arcsec.}
\label{fig_m51b_img}
\end{center}
\end{figure*}

Upon examination of the new \chandra\ data, with its improved resolution of the M51b nuclear region, we noted several X-ray point sources which could be identified as the nucleus. In order to identify the true nuclear source, we investigated data from longer wavelengths, specifically high resolution {\it HST}/WFC3 data. We present the multiwavelength images in Figure \ref{fig_m51b_img}. The nucleus is evident as a strong centrally peaked near-infrared (NIR) source, whereas in the near-ultraviolet the nuclear region is more extended. We use the position of the NIR source to inform us which is the nuclear X-ray source. Recent results from \cite{rampadarath18} using high resolution ($<1$\arcsec, $\sim$10 pc) e-MERLIN L (1--2 GHz) and C-band (4--8 GHz) radio data identify the same X-ray source as the nucleus.

\begin{figure}
\begin{center}
\includegraphics[width=90mm]{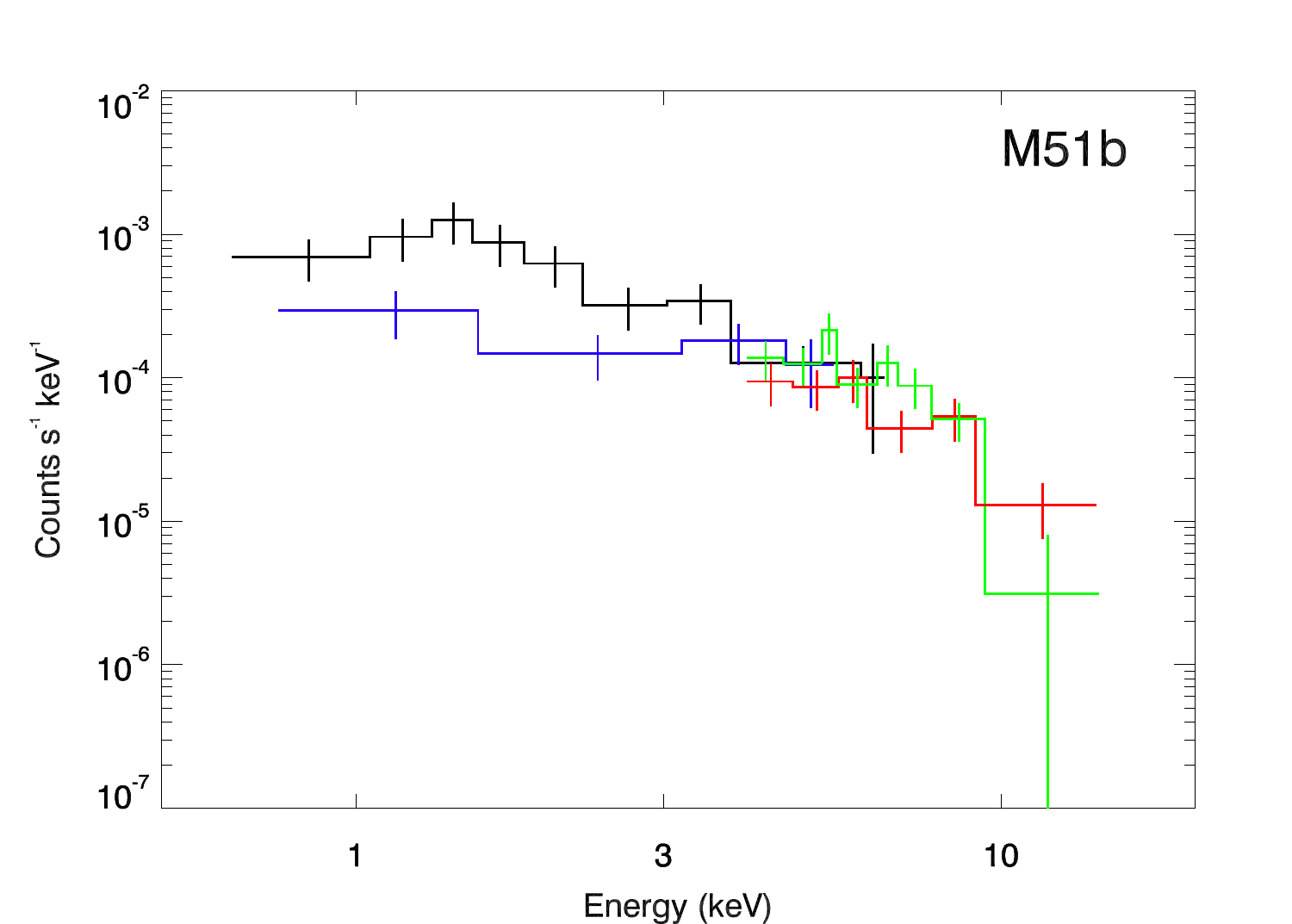}
\includegraphics[width=90mm]{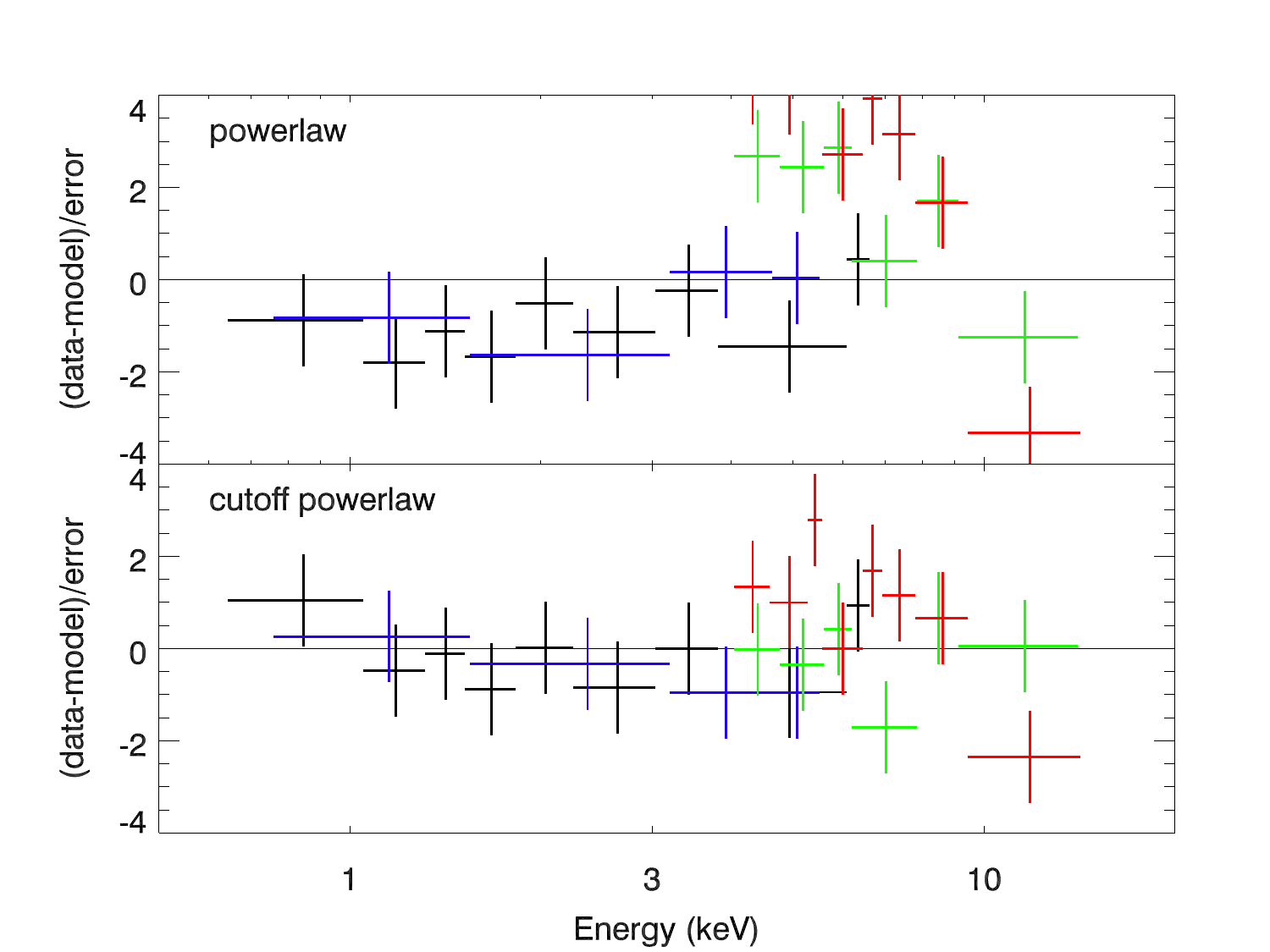}
\caption{\chandra\ (black is the nucleus and blue is the extra-nuclear source) and \nustar\ (FPMA in red and FPMB in green) spectra of the M51b nuclear region (top panel). Residuals to a fit with a power-law model for both sources, and a cut off power-law model for both are also shown (bottom panels).}
\label{fig_M51b_spec}
\end{center}
\end{figure}

We extract the \chandra\ spectra from the nucleus and the brightest extra-nuclear X-ray source within the 20\arcsec\ radius of the \nustar\ extraction region and jointly fit these along with the \nustar\ spectra. The joint spectrum is shown in Figure \ref{fig_M51b_spec}. We only consider data up to 15 keV due to the data being dominated by background above these energies.

We start with a simple power-law model for each source. We do not find evidence for absorption, with upper limits of 2$\times10^{21}$ \cmsq\ and 1$\times10^{21}$ \cmsq\ for the nucleus and extra-nuclear sources respectively. We also find an excess of soft X-rays from the extra-nuclear source that could be due to emission from a photoionized plasma. We fit it with an {\tt apec} model with the temperature fixed at 0.5 keV, in addition to the power-law model. The fit statistic for the joint \chandra\ plus \nustar\ spectrum with a simple power-law model for each source is $C/$DoF=226.4/186. The fit residuals, shown in Figure \ref{fig_M51b_spec}, show that a simple power-law model does not adequately account for the spectral shape. Allowing a high energy cut off for the extra-nuclear source improves the fit to $C/$DoF=175.5/185. Further allowing the nucleus to have a high energy cut off also improves the fit to 164.2/184. For the {\tt cutoffpl} model of the nucleus, $\Gamma=0.59^{+ 0.58}_{- 0.75}$ and $E_{\rm C}=3.3^{+  3.6}_{-  1.5}$ keV.

We list the spectral parameters for both sources in Table \ref{tab_m51b_specpar}. The intrinsic 0.5--30 keV X-ray luminosity of the AGN in M51b is 5.4$^{+  1.4}_{-  1.0}\times10^{38}$ \ergs, two orders of magnitude lower than M51a. There is no evidence for an obscured, more powerful AGN in the galaxy from the deep \nustar\ data.

\begin{table}
\centering
\caption{M51b X-ray spectral parameters}
\label{tab_m51b_specpar}
\begin{center}
\begin{tabular}{l l l l l}
\hline
$\Gamma$ & $E_{\rm C}$ & $C$/DoF & \fx\ & \lx \\
(1) & (2) & (3) & (4) & (5) \\
\hline

\multicolumn{5}{l}{\bf nucleus}\\
 0.59$^{+ 0.58}_{- 0.75}$ &  3.3$^{+  3.6}_{-  1.5}$ &\multirow{2}{*}{164.2/ 184}& 6.2$^{+ 1.7}_{- 1.2}$ &  5.4$^{+  1.4}_{-  1.0}$ \\
\multicolumn{5}{l}{\bf extra-nuclear source}\\
 $<1.6$ &  1.3$\pm0.2$ & & 4.3$^{+ 1.5}_{- 1.4}$ &  3.7$^{+  1.3}_{-  1.2}$ \\

\hline
\end{tabular}
\tablecomments{The best-fit parameters for the cut-off power-law model to the \chandra\ and \nustar\ spectrum of the nuclear sources of M51b. Column (1) gives the power-law index of the model, column (2) lists the cut-off energy of the cut-off power-law model in keV. Column (3) gives the C-statistic of the fit and the number of degrees of freedom, column (4) gives the unabsorbed flux in the range 0.5--30 keV in units of $10^{-14}$ \ergcms, and column (6) gives the luminosity assuming a distance of 8.58 Mpc to M51 in units of $10^{38}$ \ergs.}

\end{center}
\end{table}

\subsection{The ultraluminous X-ray sources}

The most notable feature in the X-ray spectra of ULXs that is not seen in the X-ray spectra of any sub-Eddington accreting black holes is a spectral turnover below 10 keV. This was first seen in high signal-to-noise observations with \xmm\  \citep[e.g.][]{stobbart06, gladstone09}. This spectral shape is generally interpreted as the superposition of one or more disk-like components. For low signal-to-noise spectra, this continuum shape can be reproduced by a simple phenomenological power-law with an exponential cut off. This is described by $F_{\gamma}=NE^{-\Gamma}e^{-E/E_C}$, where $F_{\gamma}$ is photon flux in units of photons\,keV$^{-1}$\,cm$^{-2}$\,s$^{-1}$, and $N$ is a constant with the same units. Photon energy is $E$ in keV, $\Gamma$ is the X-ray spectral index, and $E_{\rm C}$ is the cut off energy, also in keV.

Regarding disk models, the standard Shakura \& Sunyaev multicolor thin disk model has been used extesively to model the accretion disk emission from accreting black holes \citep{shakura73}. This model describes the local temperature of the disk, $T$, as proportional to radius, $r$ as $T(r)\propto r^{-p}$, where $p=\frac{3}{4}$ . However, for high accretion rate systems, a slim disk is expected \citep{abramowicz88}. For a slim disk, the local temperature of the disk has a flatter temperature profile as a function of radius with $p\sim\frac{1}{2}$ \citep{watarai00}. Slim disks have been proposed as mechanisms to explain ULXs as super-Eddington stellar remnant black hole accretors \citep[e.g.][]{kato98,poutanen07}.

In our fits of the joint \chandra\ and \nustar\ spectra we use both a cut off power-law model ({\tt cutoffpl} in {\sc xspec}) to test for the presence of a spectral turnover, and a multicolor disk model with a variable $p$ parameter to test for the emission from a slim disk ({\tt diskpbb} in {\sc xspec}).

\subsubsection{ULX3 in M51a}
\label{subsec_ulx3}
CXOU~J132950.6+471155 is located at RA=13 29 50.68, Dec=+47 11 55.2 (J2000) and named ULX3 by \cite{liu05}. The ULX is located close enough to the nucleus of M51a that \nustar\ cannot resolve it, therefore we treat it as part of the spectral fit of the nucleus, as described in Section \ref{subsec_m51a}. The spectrum of ULX3 can be described well with the cut off power-law model where $\Gamma=-2.21^{+ 0.05}_{- 0.05}$ and $E_{\rm C}=1.2^{+ 0.2}_{-0.1}$ keV. Alternatively, a fit with the {\tt diskpbb} yields an inner disk temperature of 1.58$^{+ 0.25}_{- 0.12}$ keV and a radial temperature profile index $p>0.7$. However the fit statistic is poorer than the {\tt cutoffpl} model. We show the spectra in Figures \ref{fig_M51a_spec} and \ref{fig_M51a_spec4}. Using the {\tt cflux} model in {\sc xspec} to calculate the flux of the {\tt cutoffpl} component, ULX3 has a 0.5--30 keV flux of 2.3$^{+ 0.3}_{- 0.2}\times10^{-13}$ \ergcms\ which assuming isotropic emission and a distance of 8.5 Mpc implies a luminosity of 2.0$^{+  0.3}_{-  0.2}\times10^{39}$ \ergs.

Previous \chandra\ and \xmm\ observations of this source were presented in \cite{terashima04} and \cite{dewangan05}, the authors of which referred to it as source 26. While there was possible confusion with the nucleus in the \xmm\ data, both works found the spectrum of the ULX was consistent with a power-law model, while also noting that the spectrum was very hard. Despite the \nustar\ data also containing the spectrum of the nucleus, our detailed spectral decomposition using \chandra\ has allowed us to rule out a simple power-law model for this source and has the tightest upper limit on the energy of the cut off for any ULX studied here.

\subsubsection{ULX5 in M51a}
\label{subsec_ulx5}
RX~J132954+47145 is located at RA=13 29 53.72, Dec=+47 14 35.7 (J2000) and named ULX5 by \cite{liu05}. The 0.5--30 keV spectrum of ULX5 can be described well with a simple power-law model where $\Gamma=2.23^{+ 0.34}_{- 0.31}$ and with \nh$\sim4.5\times10^{21}$ \cmsq. The fit statistic is $C=179.5$ with 200 DoFs. The inclusion of an exponential cut off improves the fit to $C=171.7$ with 199 DoFs ($\Delta C=-7.8$ for the addition of 1 free parameter). 

In order to assess if the inclusion of this parameter has improved the fit significantly, we run spectral simulations. Using the background and response files for the observed data, we simulate 5000 spectra in {\sc xspec} using the {\tt fakeit} command based on the best-fit power-law model. We then refit the simulated data in the same way as the observed data, first with the absorbed power-law then then the power-law model with a cut off, noting the improvement in $C$ each time if any. We find that only in 19 simulated spectra does the addition of a cut off lead to an improvement in $C$ of 7.8 or more. This represents a false-alarm rate of 0.4\%, which is equivalent to a $\sim3\sigma$ detection of the cut off. We therefore conclude that a spectral turnover is present in ULX5.

\begin{figure}
\begin{center}
\includegraphics[width=90mm]{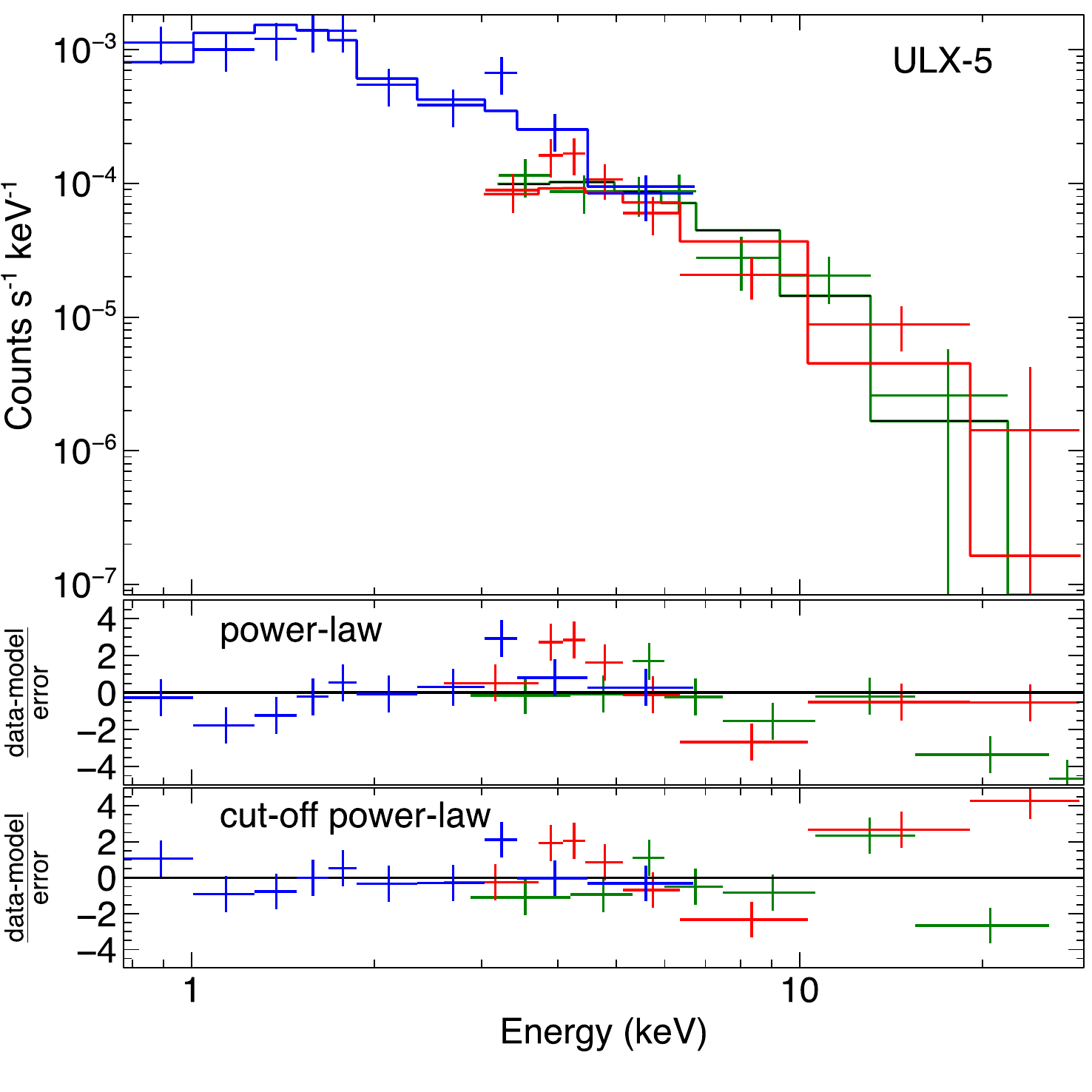}
\caption{\chandra\ (blue) and \nustar\ (FPMA in red and FPMB in green) spectra of ULX5 fitted with the cut-off power-law model are shown in the top panel. Residuals to the power-law and cut-off power-law models are shown in the bottom panels.}
\label{fig_ulx5_spec}
\end{center}
\end{figure}

For the cut off power-law model we find $\Gamma=0.89^{+ 0.65}_{- 0.81}$ and $E_{\rm cut}=3.8^{+ 14.6}_{-  1.8}$ keV. For the {\tt diskpbb} model, $T_{\rm in}=2.28^{+ 1.83}_{- 0.72}$ keV and p=0.6$^{+  0.2}_{-  0.1}$, with $C=172.9$ with 199 DoFs. We show the spectra in Figure \ref{fig_ulx5_spec} with the data to model ratios for both the power-law and cut-off power-law models.

From the cut-off power-law model ULX5 has a 0.5--30 keV flux of 1.1$^{+  0.3}_{-  0.2}\times10^{-13}$ \ergcms\ which assuming isotropic emission and a distance of 8.5 Mpc implies a luminosity of 9$^{+  3}_{-  1}\times10^{38}$ \ergs.

Results from previous \xmm\ observations of this source indicated that its spectrum was consistent with a power-law, with an added soft component that could be modeled with a multicolor disk or {\tt mekal} model \citep[][their Source 41]{dewangan05}. \cite{winter06} also studied this source using \xmm\ data, finding that it required a two component fit, with a black body and a power-law component. They measured $\Gamma=1.97$ and a flux of 2.6$\times10^{-13}$ \ergcms. 

Our sensitivity at low energies is too low to detect this extra component, however. It is also possible that this component is extended and our \chandra\ data have resolved it out. We check the location around ULX5 in our \chandra\ data and find a second point source $\sim5$ \arcsec\ to the south. The second source is not bright enough to contribute to the \nustar\ spectrum, with a 3--8 keV count rate $<10$\% of ULX5, but it does appear softer than ULX5, which may explain this second component seen in \xmm\ data.

Our data are the first to show evidence for a spectral turn-over at $\sim$10 keV in this source, which has become a hallmark of ULXs.

\subsubsection{ULX7 in M51a}
\label{subsec_ulx7}
RX J133001+47137 is located at RA+13 30 01.01, Dec=+47 13 43.9 (J2000) and was called ULX7 by \cite{liu05}. \cite{earnshaw16} studied this source in detail, noting its very high short-term variability. They found evidence for a break in the power spectrum similar to those seen in black hole binaries observed in the hard state, suggesting that this ULX is a good IMBH candidate based on these properties.

When fitting the joint \chandra\ and \nustar\ spectra of this ULX, we found evidence that a cross-calibration constant of unity was not satisfactory, and that a value $\sim3$ was required to account for the difference between the \chandra\ and \nustar\ fluxes. When fitting the spectra between the two \nustar\ observations separately, we found that the source had dropped in flux by a factor of $\sim4$ from the first observation to the second, over a period of 1--2 days. Since the \chandra\ observation overlapped mostly with the first \nustar\ observation, we use only the first observation for spectral analysis.

We then found that 0.5--30 keV spectrum of ULX7 is fitted well with a simple power-law model with $\Gamma=1.92^{+ 0.38}_{- 0.34}$ and no evidence for absorption above the Galactic column. The fit statistic was $C=110.3$ with 161 DoFs. The inclusion of an exponential cut off improves the fit statistic to 104.9 for 160 DoFs ($\Delta C=-5.4$ for the addition of 1 free parameter). Running spectral simulations in the same way as for ULX5, we find that only 41/5000 simulated spectra produce an improvement in $C$ as large or larger than we find here. This represents a false-alarm rate of 0.8\%, which is equivalent to a $\sim2.6\sigma$ detection of the cut off. We therefore conclude that a spectral turnover is present in ULX7.

\begin{figure}
\begin{center}
\includegraphics[width=90mm]{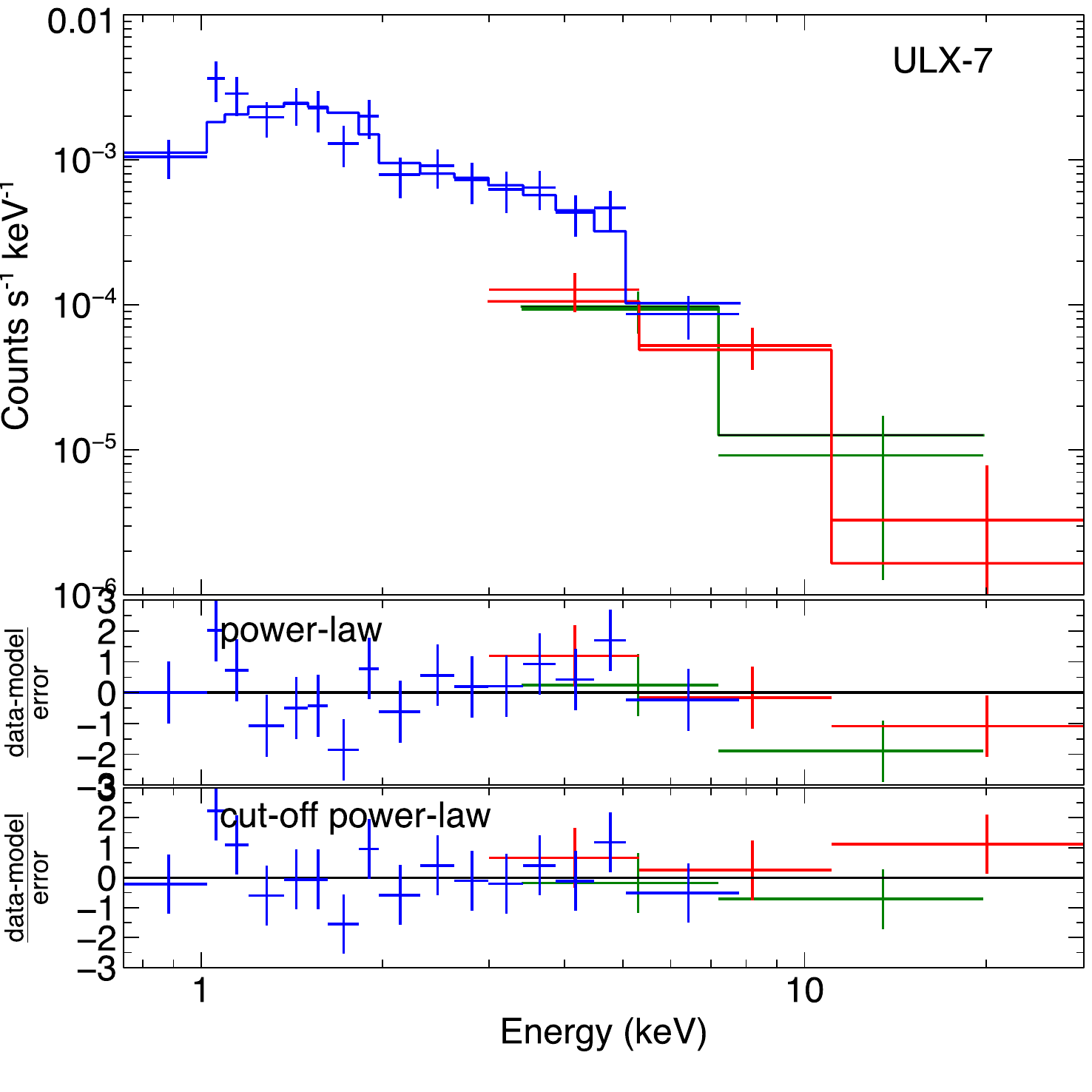}
\caption{\chandra\ (blue) and \nustar\ (FPMA in red and FPMB in green) spectra of ULX7 fitted with the cut-off power-law model are shown in the top panel. Residuals to the power-law and cut-off power-law models are shown in the bottom panels.}
\label{fig_ulx7_spec}
\end{center}
\end{figure}

For the cutoff power-law model, $\Gamma=0.60^{+ 1.27}_{- 0.90}$ and $E_{\rm C}=3.3^{+ 45.6}_{-  1.5}$ keV. For the {\tt diskpbb} model, $T_{\rm in}=2.11^{+ 6.46}_{- 0.71}$ keV and $p>0.5$, with $C=104.6$ with 160 DoFs. We show the spectra in Figure \ref{fig_ulx7_spec} with the data to model ratios for both the power-law and cut-off power-law models. During the \chandra\ observation ULX7 has a 0.5--30 keV flux of 1.2$^{+ 0.2}_{- 0.2}\times10^{-13}$ \ergcms\ which, assuming isotropic emission and a distance of 8.5 Mpc, implies a luminosity of 1.2$^{+  0.5}_{-  0.2}\times10^{39}$ \ergs.

\cite{earnshaw16} analyzed 5 \xmm\ and 11 \chandra\ observations of ULX7, and found that the source exhibits variability of over an order of magnitude in flux from a 0.3--10 keV flux of $\sim3\times10^{-14}-1\times10^{-12}$ \ergcms, but no strong spectral variability. They found a typical $\Gamma\sim1.5$ when fitting below 10 keV with a simple power-law model. When data from a short \nustar\ observation were included, they showed that inclusion of a cut off in power-law spectrum improved their \chisq\ statistic by 5, finding $\Gamma=1.3\pm0.1$ and $E_{\rm C}=18^{+43}_{-8}$ keV. This is broadly consistent with the results we find, within the large uncertainties.

Since \cite{earnshaw16} did not find evidence for spectral variability, we investigated using all the \nustar\ data from the new observations rather than just those overlapping with the \chandra\ data as described above. We accounted for the drop in flux from the source with a variable cross-calibration constant. However, since the source had dropped in flux, using the entire observation, rather than just the segment at high flux, did not improve the counting statistics significantly, and no new constraints could be placed on the parameters, finding $\Gamma=0.69^{+ 1.27}_{- 0.90}$ and $E_{\rm C}=3.7^{+u}_{-  1.4}$ keV (i.e. unconstrained at the upper end) where $C=185.0$ with 220 DoFs for the {\tt cutoffpl} model.

\subsubsection{ULX8 in M51a}
\label{subsec_ulx8}

RX J133007+47110 is located at RA=13 30 07.55, Dec=+47 11 06.1 (J2000) and was named ULX8 by \cite{liu05} or NGC 5194 X8/ULX5 by \cite{liu05b}. This was recently found to be powered by a neutron-star accretor, implied from the detection of a cyclotron resonance scattering feature (CRSF) in an archival 2012 \chandra\ observation \citep{brightman18}. During the 2012 \chandra\ observation the source was observed flaring to fluxes up to 10$^{-12}$ \ergcms, or luminosities up to $10^{40}$ \ergs, from more typical fluxes of $\sim10^{-13}$ \ergcms. The CRSF was observed at 4.5 keV with a Gaussian width of 0.1 keV and an equivalent width of -0.19$^{+0.06}_{-0.09}$ keV. Furthermore, the high signal-to-noise \chandra\ spectrum showed a significant departure from a simple power-law spectrum. The continuum could be fitted by a power-law model ($\Gamma=1.3\pm0.3$) with an exponential cut off ($E_{\rm C}=3.7^{+2.2}_{-1.0}$ keV).

\begin{figure}
\begin{center}
\includegraphics[width=90mm]{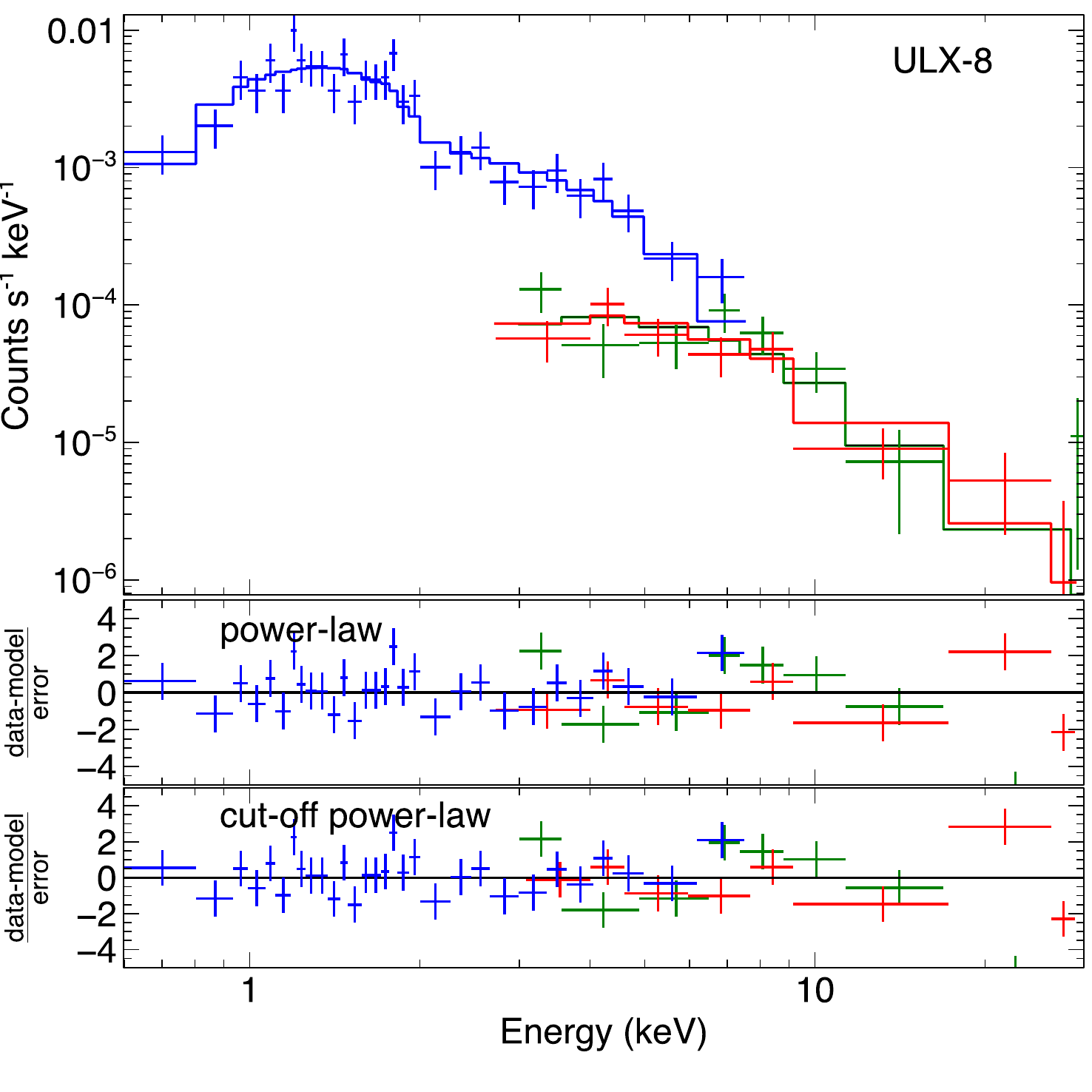}
\caption{\chandra\ (blue) and \nustar\ (FPMA in red and FPMB in green) spectra of ULX8 fitted with the power-law model are shown in the top panel. Residuals to the power-law and cut-off power-law models are shown in the bottom panels.}
\label{fig_ulx8_spec}
\end{center}
\end{figure}

During our new \chandra\ and \nustar\ observations, the source was observed at a much lower 0.5--30 keV flux of 1.6$^{+  0.2}_{-  0.2}\times10^{-13}$ \ergcms. The 0.5--30 keV spectrum can be described well with a simple power-law model with $\Gamma=1.94^{+ 0.23}_{- 0.20}$ where $C=228.7$ with 304 degrees of freedom. The inclusion of an exponential cut off does not improve the fit statistic significantly ($\Delta C=-0.2$ for the addition of 1 free parameter) and the parameters of the {\tt diskpbb} model are not constrained by the data. We show the spectra in Figure \ref{fig_ulx8_spec} with the data to model ratios for both the power-law and cut-off power-law models.

We furthermore find no evidence for absorption lines in the new data. Adding an absorption line at 4.5 keV does not improve the fit statistic. However since our new observations found this source at a lower flux than the 2012 \chandra\ observation, the CRSF is not likely to be detectable. The 90\% confidence lower limit on the equivalent width of an absorption line at 4.5 keV is  -0.18 keV, which is consistent with that measured in the 2012 \chandra\ data. The shape of the X-ray spectrum measured here is marginally consistent with that measured in the 2012 \chandra\ data, but we cannot rule out spectral evolution with flux.

\cite{dewangan05} also found that this source (their source \#82) was consistent with an absorbed power-law from its \xmm\ spectrum, with $\Gamma=2.4\pm0.2$ observed at a slightly higher flux of 2.6$\times10^{-13}$ \ergcms. They also found higher absorption than evident in our \chandra\ spectrum with \nh$=1.6\pm0.4\times10^{21}$ \cmsq. \cite{yoshida10} analyzed 3 \chandra\ and 4 \xmm\ observations of this source, again finding that its spectrum is consistent with a power-law.

\subsubsection{ULX9 in M51b}
\label{subsec_ulx9}

RX J133006+47156 is located at RA+13 30 06.00, Dec=+47 15 42.3 (J2000) and was called ULX9 by \cite{liu05}. A fit with an absorbed power-law model to the 0.5--30 keV spectrum of ULX9 gives $\Gamma=2.34^{+ 0.33}_{- 0.30}$ and \nh$\sim7.8\times10^{21}$ \cmsq. The fit statistic is $C=271.5$ with 243 degrees of freedom. Including an exponential cut off improves the fit to $C=258.1$ with 242 degrees of freedom ($\Delta C=-13.4$ for the addition of 1 free parameter). Spectral simulations  show that only 1/5000 simulated spectra produce an improvement in $C$ as large or larger than we find here. This represents a false-alarm rate of $\sim$0.02\%, which is equivalent to a $>3\sigma$ detection of the cut off. We therefore conclude that a spectral turnover is present in ULX9.

\begin{figure}
\begin{center}
\includegraphics[width=90mm]{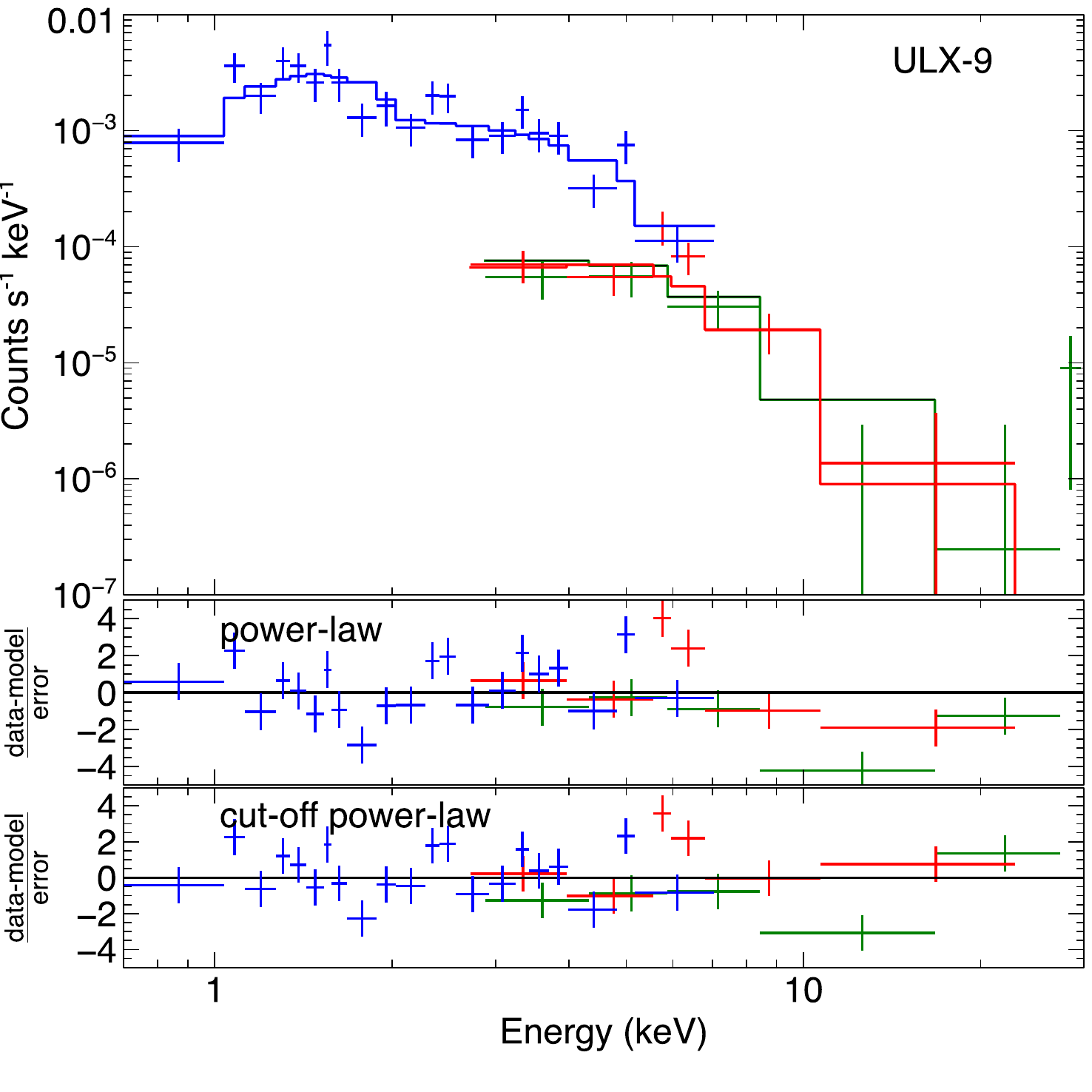}
\caption{\chandra\ (blue) and \nustar\ (FPMA in red and FPMB in green) spectra of ULX9 fitted with the cut-off power-law model are shown in the top panel. Residuals to the power-law and cut-off power-law models are shown in the bottom panels.}
\label{fig_ulx9_spec}
\end{center}
\end{figure}

We find $\Gamma=0.35^{+ 1.26}_{- 1.05}$ and $E_{\rm C}=2.3^{+  4.2}_{-  1.3}$ keV for the {\tt cutoffpl} model. For the {\tt diskpbb} model, $T_{\rm in}=1.82^{+ 1.21}_{- 0.47}$ keV and p is unconstrained. Here $C=257.7$ with 242 DoFs. From the cut off power-law model ULX9 has a 0.5--30 keV flux of 9$^{+  2}_{-  1}\times10^{-14}$ \ergcms\ which assuming isotropic emission and a distance of 8.5 Mpc implies a luminosity of 8$^{+  2}_{-  1}\times10^{38}$ \ergs. We show the spectra in Figure \ref{fig_ulx9_spec} with the data to model ratios for both the power-law and cut-off power-law models.

Previous modeling of the \xmm\ spectrum found that a power-law alone could explain the shape \citep{dewangan05}, but at a higher flux of 1.2$\times10^{-13}$ \ergcms.

\subsubsection{J132946+471041}
\label{subsec_2946}
\begin{figure}
\begin{center}
\includegraphics[width=90mm]{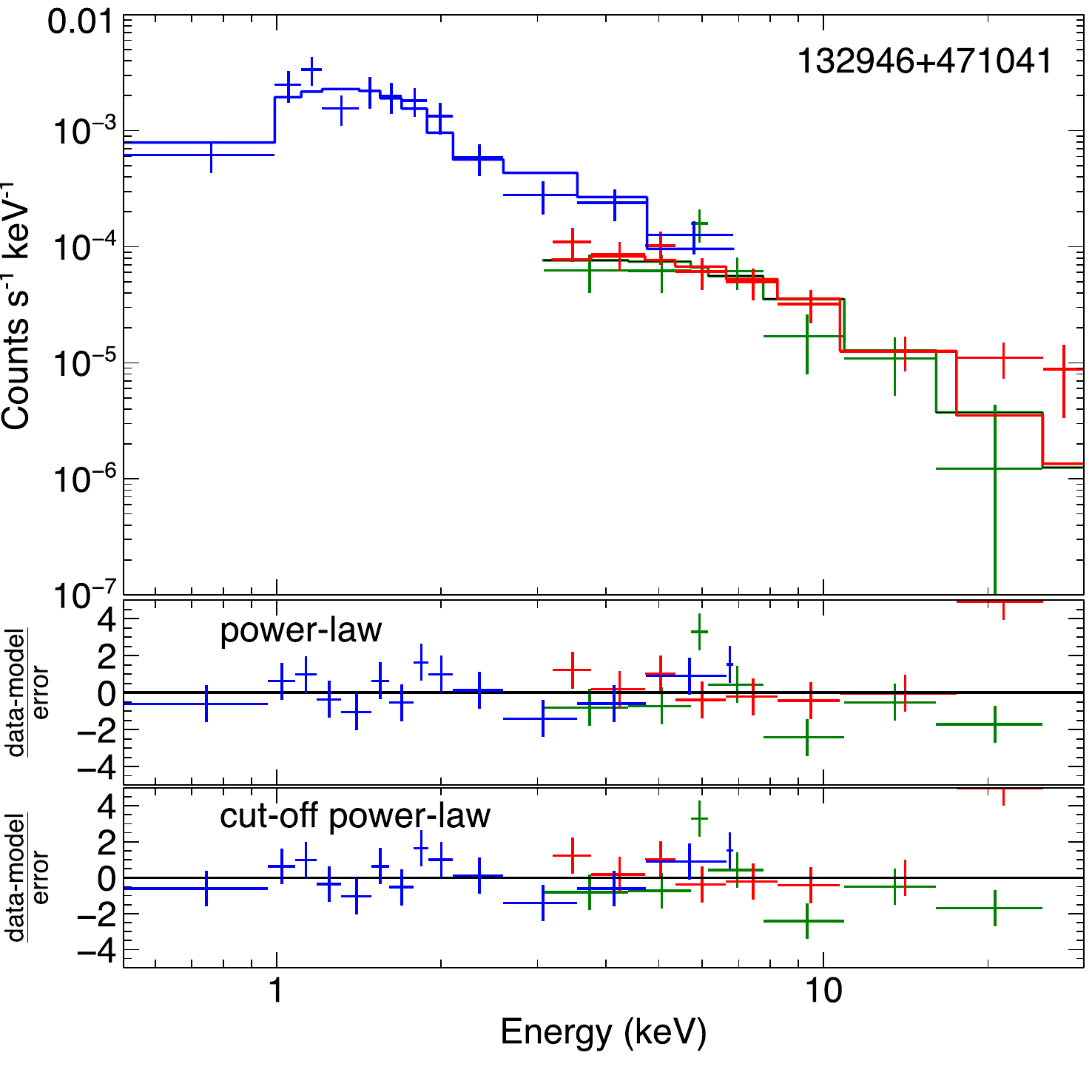}
\caption{\chandra\ (blue) and \nustar\ (FPMA in red and FPMB in green) spectra of J132946+471041 fitted with the power-law model are shown in the top panel. Residuals to the power-law and cut-off power-law models are shown in the bottom panels.}
\label{fig_2946_spec}
\end{center}
\end{figure}

RX J132946+47107 is located at RA=13 29 46.11, Dec=+47 10 42.3 (J2000). The 0.5--30 keV spectrum of this source is described well by a power-law model with $\Gamma=1.64^{+ 0.26}_{- 0.16}$, no absorption above the Galactic \nh\ and no evidence for a cut off. The parameters of the {\tt diskpbb} model are unconstrained. The 0.5--30 keV flux is 1.1$\times10^{-13}$ \ergcms\ implying a luminosity of 9$^{+  2}_{-  3}\times10^{38}$ \ergs. We show the spectra in Figure \ref{fig_2946_spec} with the data to model ratios for both the power-law and cut-off power-law models.

\subsubsection{J132959+471052}
\label{subsec_2959}

\begin{figure}
\begin{center}
\includegraphics[width=90mm]{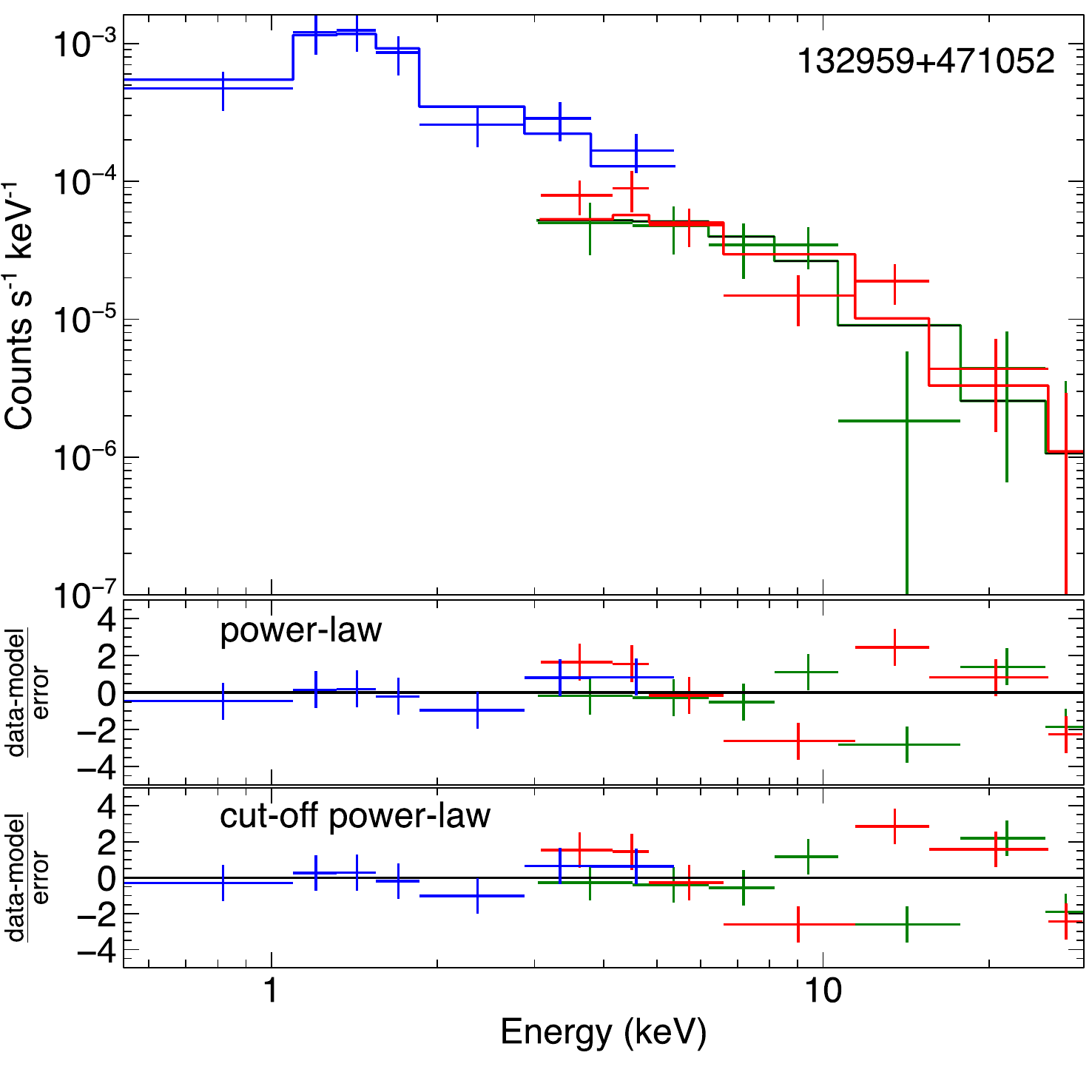}
\caption{\chandra\ (blue) and \nustar\ (FPMA in red and FPMB in green) spectra of J132959+471052 fitted with the power-law model are shown in the top panel. Residuals to the power-law and cut-off power-law models are shown in the bottom panels.}
\label{fig_2959_spec}
\end{center}
\end{figure}

CXOU J132957.5+471048 is located at RA=13 29 57.57, Dec=+47 10 48.3 (J2000) and was called XMM6 by \cite{winter06}. The 0.5--30 keV spectrum of this source is described well by a power-law model with $\Gamma=1.33^{+ 0.28}_{- 0.21}$, no absorption above the Galactic \nh\ and no evidence for a cut off. The 0.5--30 keV flux is 7.6$\times10^{-14}$ \ergcms\ implying a luminosity of 7$\times10^{38}$ \ergs. We show the spectra in Figure \ref{fig_2959_spec} with the data to model ratios for both the power-law and cut-off power-law models.

\begin{table*}
\centering
\caption{Extra-nuclear point-source X-ray spectral parameters}
\label{tab_specpar}
\begin{center}

\begin{tabular}{l l l l l l l}
\hline
Name and model &\nh & $\Gamma$ or $T_{\rm in}$ & $E_{\rm C}$ or $p$ & $C$/DoF & \fx\ & \lx \\
(1) & (2) & (3) & (4) & (5) & (6)  & (7) \\

\hline

\multicolumn{6}{l}{\bf ULX3}\\
{\tt cutoffpl} & 0.0 & -2.21$^{+ 0.05}_{- 0.05}$ &  1.2$^{+ 0.2}_{-  0.1}$ &315.7/ 258& 2.3$^{+ 0.3}_{- 0.2}$ &  2.0$^{+  0.3}_{-  0.2}$ \\
{\tt diskpbb} &  0.0 & 1.58$^{+ 0.25}_{- 0.12}$ &  $>0.7$ &322.4/ 258 &  2.7$^{+ 0.3}_{- 0.2}$ &  2.4$^{+  0.2}_{-  0.2}$ \\
\multicolumn{7}{l}{\bf ULX-5}\\
{\tt powerlaw} &  4.5$^{+  4.2}_{-  3.6}$ & 2.23$^{+ 0.34}_{- 0.31}$ & - &179.5/ 200&  1.5$^{+  0.3}_{-  0.2}$ &  1.3$^{+  0.3}_{-  0.2}$ \\
{\tt cutoffpl} &  $<4.7$ & 0.89$^{+ 0.65}_{- 0.81}$ &  3.8$^{+ 14.6}_{-  1.8}$ &171.7/ 199&  1.1$^{+  0.3}_{-  0.2}$ &  0.9$^{+  0.3}_{-  0.1}$ \\
{\tt diskpbb} &  0.0 & 2.28$^{+ 1.83}_{- 0.72}$ &  0.6$^{+  0.2}_{-  0.1}$ &172.9/ 199&  1.1$^{+  0.2}_{-  0.2}$ &  0.9$^{+  0.2}_{-  0.1}$ \\
\multicolumn{7}{l}{\bf ULX-7}\\
{\tt powerlaw} &  $<3.2$ & 1.92$^{+ 0.38}_{- 0.34}$ & - &110.3/ 161&  1.8$^{+  0.5}_{-  0.3}$ &  1.6$^{+  0.4}_{-  0.3}$ \\
{\tt cutoffpl} &  $<3.1$ & 0.60$^{+ 1.27}_{- 0.90}$ &  3.3$^{+ 45.6}_{-  1.5}$ &104.9/ 160&1.2$^{+ 0.2}_{- 0.2}$ &  1.2$^{+  0.5}_{-  0.2}$ \\
{\tt diskpbb} &  $<4.5$ & 2.11$^{+ 6.46}_{- 0.71}$ &  $>0.5$ &104.6/ 160&  1.4$^{+  0.5}_{-  0.2}$ &  1.2$^{+  0.4}_{-  0.2}$ \\
\multicolumn{7}{l}{\bf ULX-8}\\
{\tt powerlaw} &  $<0.9$ & 1.94$^{+ 0.23}_{- 0.20}$ & - &228.7/ 304&  1.6$^{+  0.2}_{-  0.2}$ &  1.4$^{+  0.2}_{-  0.2}$ \\
{\tt cutoffpl} &  $<0.6$ & 1.82$^{+ 0.24}_{- 0.50}$ & 41.4$^{+u}_{- 36.5}$ &228.5/ 303&  1.5$^{+  0.3}_{-  0.2}$ &  1.3$^{+  0.2}_{-  0.2}$ \\
{\tt diskpbb} &  0.0 & 6.41$^{+u}_{- 3.39}$ &  0.5$^{+  u}_{- l}$ &228.9/ 303&  1.5$^{+  0.2}_{-  0.2}$ &  1.3$^{+  0.2}_{-  0.2}$ \\
\multicolumn{7}{l}{\bf ULX-9}\\
{\tt powerlaw} &  7.8$^{+  3.7}_{-  3.2}$ & 2.34$^{+ 0.33}_{- 0.30}$ & - &271.5/ 243&  1.4$^{+  0.3}_{-  0.2}$ &  1.2$^{+  0.2}_{-  0.2}$ \\
{\tt cutoffpl} &  $<1.5$ & 0.35$^{+ 1.26}_{- 1.06}$ &  2.3$^{+  4.2}_{-  1.3}$ &258.1/ 242&  0.9$^{+  0.2}_{-  0.1}$ &  0.8$^{+  0.2}_{-  0.1}$ \\
{\tt diskpbb} &  $<2.3$ & 1.82$^{+ 1.21}_{- 0.47}$ &  0.7$^{+u}_{-  l}$ &257.7/ 242&  0.9$^{+  0.2}_{-  0.1}$ &  0.8$^{+  0.2}_{-  0.1}$ \\
\multicolumn{7}{l}{\bf 132946+471041}\\
{\tt powerlaw} &  $<0.1$ & 1.64$^{+ 0.26}_{- 0.16}$ & - &221.8/ 303&  1.1$^{+  0.2}_{-  0.2}$ &  0.9$^{+  0.2}_{-  0.2}$ \\
{\tt cutoffpl} &  $<0.1$ & 1.62$^{+ 0.26}_{- 0.43}$ &500.0$^{+u}_{-l}$ &219.5/ 223&  1.1$^{+  0.2}_{-  0.3}$ &  0.9$^{+  0.2}_{-  0.3}$ \\
{\tt diskpbb} &  $<2.7$ &10.00$^{+u}_{- 5.48}$ &  0.6$^{+  u}_{-  l}$ &220.4/ 223&  1.0$^{+  0.2}_{-  0.2}$ &  0.8$^{+  0.2}_{-  0.2}$ \\
\multicolumn{7}{l}{\bf 132959+471052}\\
{\tt powerlaw} &  $<3.0$ & 1.33$^{+ 0.28}_{- 0.21}$ & - &198.8/ 177&  0.8$^{+  0.3}_{-  0.2}$ &  0.7$^{+  0.3}_{-  0.2}$ \\
{\tt cutoffpl} &  0.0 & 1.13$^{+ 0.34}_{- 0.66}$ & 23.8$^{+u}_{- l}$ &198.2/ 176&  0.6$^{+  0.2}_{-  0.1}$ &  0.5$^{+  0.1}_{-  0.1}$ \\
{\tt diskpbb} &  $<2.6$ & 7.98$^{+u}_{- 4.67}$ &  0.6$^{+  0.1}_{-  0.1}$ &198.2/ 176&  0.6$^{+  0.2}_{-  0.1}$ &  0.5$^{+  0.1}_{-  0.1}$ \\

\hline
\end{tabular}
\tablecomments{Spectral parameters of the power-law, cut-off power-law and diskpbb model fits to the \chandra\ and \nustar\ spectra of the extra-nuclear point sources. For ULX3, this is part of the dataset on the nucleus. Column (2) gives the absorption column intrinsic to the source in units of $10^{21}$ \cmsq. `0.0' indicates that no absorption was detected on top of the Galactic absorption. Column (3) gives the power-law index of the power-law and cut-off power-law models or the temperature of the diskpbb model in keV, column (4) lists the cut-off energy of the cut-off power-law model in keV or the index of the radial temperature profile of the diskpbb model. `u' and/or `l' indicate that the parameter was unconstrained at the upper or lower end. Column (5) gives the C-statistic of the fit and the number of degrees of freedom, column (6) gives the unabsorbed flux in the range 0.5--30 keV in units of $10^{-13}$ \ergcms, and column (7) gives the luminosity assuming a distance of 8.58 Mpc to M51 in units of $10^{39}$ \ergs.}

\end{center}
\end{table*}

\section{discussion}

\subsection{The dual active galactic nuclei}
\label{sec_disc_agn}

Dual AGN occur when a pair of galaxies separated on kiloparsec scales are simultaneously both observed to host AGN \citep[e.g. NGC 6240,][]{komossa03} and are predicted to occur by merger-driven AGN models. The dual AGN of M51 are only the second to be resolved above 10 keV with \nustar\ after MCG +04-48-002 and NGC 6921 \citep{koss16a}. They were only recently detected by \swiftbat, but their individual hard X-ray emission could not be resolved \citep{oh18}. Prior to \nustar\ and \swiftbat, the nucleus of M51a has been studied at hard X-ray wavelengths most notably by \cite{fukazawa01} where a {\it BeppoSAX} observation of the galaxy was reported. The authors inferred a column density of $\sim10^{24}$ \cmsq\ for the nucleus based on the excess of hard emission over that seen at softer energies. An analysis of a more recent $\sim20$ ks \nustar\ observation of M51 in conjunction with deep archival \chandra\ data was presented in \cite{xu16}, where the intrinsic \lx, $\Gamma$ and \nh\ were estimated. However, the low signal to noise of the \nustar\ data did not allow measurement of the torus covering factor, which requires good quality data above 10 keV. Furthermore, the \nustar\ observation lacked the simultaneous \chandra\ data we obtained to resolve out the variable extra-nuclear emission. \cite{xu16} inferred \lx$=4\times10^{40}$ \ergs, $\Gamma=1.8\pm0.3$ and \nh$=7\pm3\times10^{24}$ \cmsq\ with the {\tt mytorus} model. For that model we obtain \lx=1.8$^{+  1.0}_{-  0.7}\times10^{40}$ \ergs, $\Gamma=1.58^{+ 0.20}_{-l}$ and \nh$=7.1^{+u}_{-  2.4}\times10^{24}$ \cmsq. The \nh\ values are in good agreement, but our luminosity estimate is slightly lower, and our estimate of the intrinsic photon index is harder.

\cite{gandhi09} noted that there exists a very tight relationship between the intrinsic X-ray luminosity and the 12 micron luminosity of AGN where the nucleus has been resolved in the mid-infrared and a good estimate of the intrinsic X-ray luminosity exists. From the latest relationship published by \cite{asmus15}, given an intrinsic X-ray luminosity of $2\times10^{40}$ \ergs, the predicted 12 micron luminosity is $5\times10^{40}$ \ergs. This is exactly the value observed for M51a from high-resolution MIR imaging, and thus adds support to this relationship down to the lowest luminosities observed of $\sim10^{40}$ \ergs.

We calculated the black hole mass of M51a using the \mbh-$\sigma_{*}$ relation from \cite{kormendy13} where log($\frac{M_{\rm BH}}{M_{\odot}}) = 4.38\times$log$(\frac{\sigma_{*}}{200\,\kms}) + 8.49$.  We used the $\sigma_{*}$ inferred from the Ca\,{\sc ii} triplet measurement of 63$\pm$4 \kms\ to determine a black hole mass as the 3950\AA\ to 5500\AA\ region velocity dispersion was close to the instrumental resolution. This yielded log(\mbh/\msol)=$6.3\pm0.4$, where the uncertainty has been propagated from the measurement uncertainty on the stellar dispersion which is larger than the intrinsic scatter of the relationship of 0.29 \citep{kormendy13}. The mass of the black hole in the nucleus of M51a was previously estimated to be log(\mbh/\msol)=6.95 by \cite{woo02} where a stellar dispersion value of 102 \kms\ from \cite{nelson95} was used. Our stellar dispersion measurement is smaller due to the better spectral resolution of our measurement. We note that \cite{ho09} also measured a lower velocity dispersion of 76.3$\pm9.1$ \kms\ for M51a, also at Palomar.

We calculate the bolometric luminosity, and subsequently the Eddington ratio of M51a, by applying a bolometric correction, $\kappa_{\rm Bol}$, to the X-ray luminosity. The X-ray luminosity of $\sim10^{40}$ \ergs\ is lower than studies of $\kappa_{\rm Bol}$ have used \citep[e.g.][]{marconi04,lusso12}. These have found a decreasing trend of $\kappa_{\rm Bol}$ with luminosity, with $\kappa_{\rm Bol}=$10 shown to be appropriate for low-luminosity AGN, including Compton-thick ones \citep{brightman17}. Using $\kappa_{\rm Bol}=$10 implies M51a has a bolometric luminosity of $\sim10^{41}$ \ergs\ and an Eddington ratio of \lamedd$\sim10^{-4}$.

The low measured value of the photon index of 1.4--1.8 (depending on the model used) for M51a is consistent with a low Eddington rate system \citep[e.g.][]{shemmer06,brightman13,trakhtenbrot17}, even when modeling the spectra of Compton-thick AGN with torus models \citep{brightman16}. For \lamedd$\sim10^{-4}$, the $\Gamma$-\lamedd\ relationship predicts a range in $\Gamma$ of 1.2--1.5 from the fit to the full BASS sample presented in \cite{trakhtenbrot17}, and therefore our results are fully consistent with this relationship.

The luminosity and Eddington ratio regime of M51a is an extremely low one, being the lowest luminosity Compton-thick AGN known, slightly less luminous than the recently identified low-luminosity CTAGN in NGC 1448 \citep{annuar17}. This allows us to test various models of torus formation in a regime where the torus is predicted to disappear or be diminished. Based on a model where the torus is produced by outflowing material, \cite{elitzur06} suggested that the obscuring torus disappears below a bolometric luminosity of $10^{42}$ \ergs. However, the mere detection of a Compton-thick line of sight to the AGN shows that this is not the case. Furthermore, the torus covering factor for the AGN in M51a was inferred to be 0.26$^{+ 0.02}_{- 0.02}$ from the {\tt borus} model, showing that while the torus subtends a small fraction of the sky, it is a significant one. The radiation driven fountain model of \cite{wada15} also predicts a diminished torus at low luminosities, or more specifically Eddington ratios, however their model still predicts a covering factor of 0.1--0.3 at the low end of their \lx\ range, having peaked at higher \lx, around $10^{43}-10^{44}$ \ergs. Their calculations do not consider X-ray luminosities as low as that observed from M51a. 

On the other hand, the fact that M51a is in an on-going merger with M51b may be more pertinent regarding the source of the obscuration, since gas will have been driven into the the nucleus during this merger process. Indeed, recent results on mergers show that AGN in these systems show increased incidence of Compton-thick obscuration with respect to those not in mergers \citep{kocevski15,ricci17a}. This may explain the presence of a large amount of obscuring material in the nucleus of M51a despite its low accretion rate and luminosity. The fact that a prominent reflection component is observed in the X-ray spectrum of M51a suggests that the obscuration is being carried out by a torus-like structure. This is because the X-ray source needs to be both reflected by Compton-thick material out of the line of sight, and obscured by material in the line of sight to produce such a feature. \chandra\ resolves this emission at $\sim1$\arcsec\ ($\sim40$ pc at 8.58 Mpc) scales confirming that the obscuring material is very close to the nucleus and not on galactic scales. The fact that the X-ray and MIR luminosities of M51a lie on the same relationship as other local AGN also implies that this material is on the same scales as the obscuring torus.

\begin{figure}
\begin{center}
\includegraphics[width=90mm]{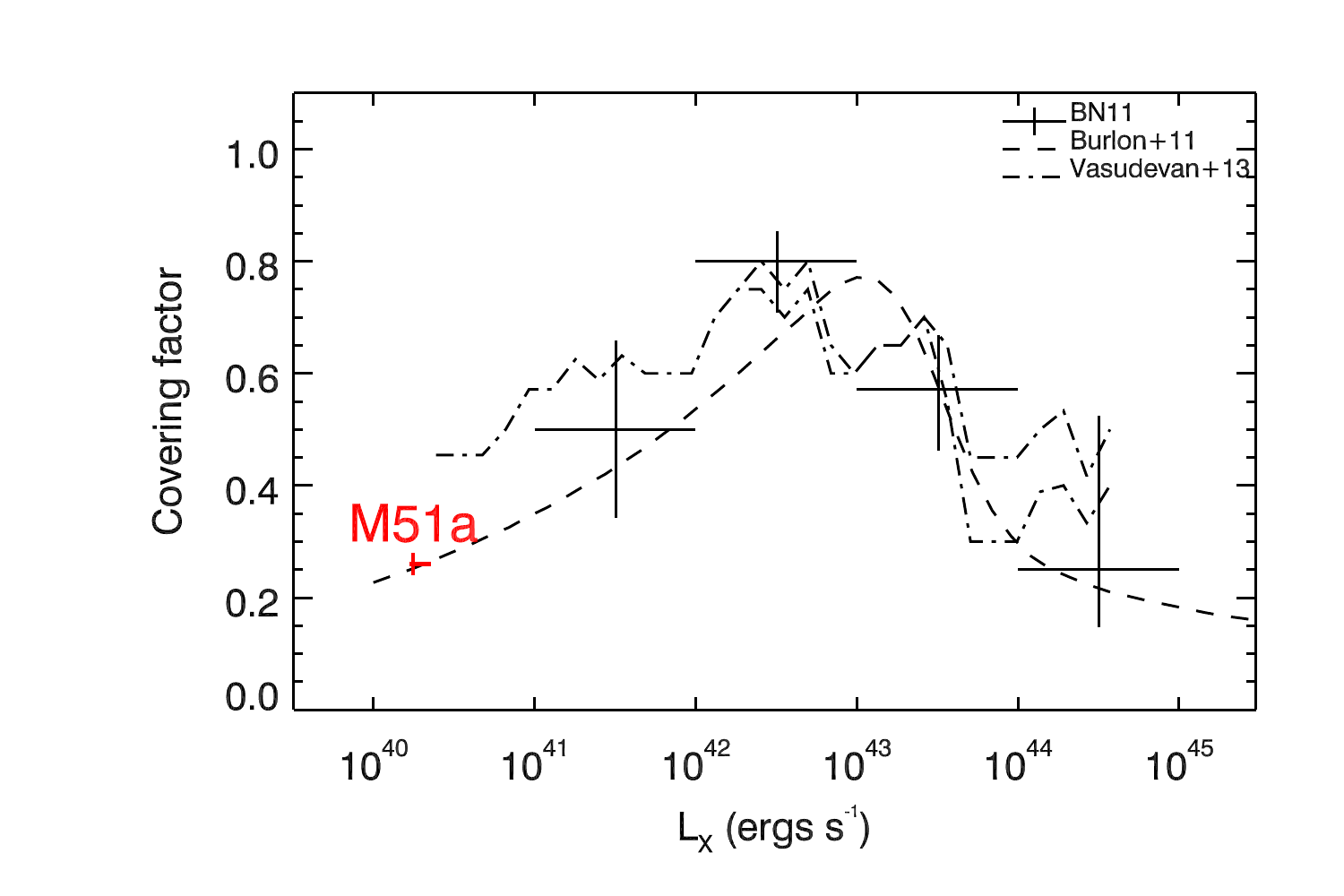}
\caption{The covering factor of the torus in the AGN of M51a derived from the {\tt borus} model, plotted in red. We compare this covering factor to the local AGN obscured fraction as derived by \cite{burlon11}, \cite{brightman11b} and \cite{vasudevan13}.}
\label{fig_M51a_fobs}
\end{center}
\end{figure}

We compare the covering factor that we have derived to the local AGN obscured fraction as a function of X-ray luminosity, which is the average torus covering factor, as derived by \cite{burlon11}, \cite{brightman11b} and \cite{vasudevan13}, and show the comparison in Figure \ref{fig_M51a_fobs}. The results agree very well, showing that M51a supports the decline in the torus covering factor at low X-ray luminosities. Previous results on inferring the covering factor of the torus in Compton-thick AGN at higher luminosities have also shown good agreement with the obscured fraction \citep[e.g.][]{brightman15}. The luminosity dependence of the obscured fraction, which has often been tied to the increasing dust sublimation radius with luminosity, has more recently been attributed to an accretion rate dependence \citep[e.g.][]{ricci17}. \cite{ricci17} found from a large sample of local \swiftbat\ detected AGN that the obscured fraction shows a sharp drop above \lamedd$\sim0.01$, which corresponds to the effective Eddington limit of dusty gas. 

While it was previously concluded that the torus in M51a must have a large covering factor in order to account for the high EW of its \feka\ line \citep{levenson02}, our new data combined with the latest {\tt borus} model find that the covering factor is relatively low. The latest calculations presented in \cite{balokovic18} show that the high EW of 3.3 keV can be produced even for low covering factors, especially when the line of sight \nh\ is high, as it is for M51a, also dependent on viewing perspective. The high EW of the \feka\ line given the low luminosity of M51a is also consistent with the X-ray Baldwin effect, otherwise known as the Iwasawa-Taniguchi effect \citep{iwasawa93}, which describes an anti-correlation between the \feka\ EW and the intrinsic X-ray luminosity. This was recently explored for a sample of CTAGN in \cite{boorman18}, finding that the relationship holds even for these sources and may be due to the luminosity-dependent torus covering factor \citep[e.g.][]{ricci13}.

For both the {\tt borus} and {\tt mytorus} models, we found that allowing the scattered and transmitted components to be decoupled from one another leads to an improvement in the fit statistic, implying that the smooth toroidal geometries that these models describe are too simplistic. This is likely due to the fact that the torus is clumpy, rather than smooth as described by the simplest unified scheme.

Several works have estimated the intrinsic X-ray luminosity of the AGN in M51b \citep[e.g.][]{hgarcia14}, however, due to its low luminosity, several bright non-nuclear sources have made this challenging, even for \chandra\ observations where M51b was off axis. With our latest observation, where M51b was closer to the optical axis than previous observations, we have resolved the nuclear region finding that its true luminosity is even lower than previous estimates with \lx=5$\times10^{38}$ \ergs. This was similarly found by \cite{rampadarath18} using the same \chandra\ data that we use. Annuar et al. (in preparation) also analyze these data as part of an investigation into the \nh\ distribution of AGN within 15 Mpc. Their \lx\ estimate for M51b is also in agreement with ours. 

We calculated the black hole mass of M51b in the same way as M51a as above using $\sigma_{*}=124.8\pm8.1$ \kms\ from \cite{ho09}. This yielded log(\mbh/\msol)=$7.6\pm0.5$. This is consistent with the value calculated by \cite{schlegel16} of 7.6. It may be surprising that the mass of the SMBH in M51b is more massive than that in M51a since M51b is often named a dwarf galaxy. However, M51b has a total stellar mass of $2.5\times10^{10}$ \msol, which is more than half the stellar mass of M51a with $4.7\times10^{10}$ \msol\ \citep{mentuch12}. Both galaxies are consistent with the distribution of \mbh\ and $M_{*}$ presented in \citep[e.g.][]{reines15}. M51b nevertheless has a higher \mbh/$M_{*}$ ratio than the average and and M51a has a lower one.

Given its X-ray luminosity and a bolometric correction of 10 implies the AGN in M51b has a bolometric luminosity of $5\times10^{39}$ \ergs\ and therefore an Eddington ratio of \lamedd$\sim10^{-6}$. M51b also has a detection by {\it Spitzer/IRS} of the mid-IR line \nev\ at 14.3\,$\mu$m which due to its high ionization potential, is considered a good tracer of AGN activity \citep{goulding09}. \cite{gruppioni16} calibrated a relationship between \lbol, derived from IR torus modelling, and the \nev\ line luminosity. For our derived \lbol\ value, the implied \nev\ line luminosity is $2.8\times10^{37}$ \ergs. The observed flux of the \nev\ line is 2$\times10^{-15}$ \ergcms\ \citep{goulding09}, which corresponds to a luminosity of $1.8\times10^{37}$ \ergs\ assuming a distance of 8.58 Mpc, in very good agreement with the \lbol\ estimate.

We also find evidence for a break in the X-ray spectrum of M51b at 2--7 keV, however since there may still be unresolved X-ray binaries in the nucleus of M51b that could mimic this spectral shape, we cannot draw strong conclusions about this pertaining to the nature of the X-ray emission from the nucleus.

The luminosities and/or accretion rates inferred for the accreting SMBHs in the M51 system are lower than predicted by galaxy merger simulations, especially considering M51 is rich in molecular gas \citep[$2\times10^{9}$\,\msol, e.g.][]{schuster07,schirm17}. \cite{vanwassenhove12} predict that for a 1:2 spiral-spiral merger, both AGN should exhibit bolometric luminosities of $>10^{41}$ \ergs\ for the entire merger period. Even for a 1:2 elliptical-spiral merger, the secondary galaxy (in this case M51b) should exhibit \lbol$\sim10^{42}$ \ergs\ or above for the entire merger. This is clearly not the case for M51.

Dual AGN activity occurs mainly at small separations \citep[$<$10 kpc, e.g.][]{satyapal14,fu18}, following the second and subsequent pericenter passages. Simulations have found that M51b has passed through M51a at least twice \citep{salo00}. The angular separation between the nuclei of the two galaxies is 4.4 arcmin, which corresponds to a projected separation of 11 kpc at 8.58 Mpc, although M51b is thought to be behind M51a, so the actual separation may be larger. Indeed dynamical modeling suggests the pericenter passage was $\sim25$ kpc \citep{salo00}, which if this corresponds to the second pericenter passage implies the M51 is in a relatively early stage of its merger compared to other systems modeled \citep[e.g.][]{hibbard95,privon13}. Since dual AGN with closer separations have higher Eddington ratios, the low Eddington ratios that we derive support that we are observing the early stages of the merger. Also, there is likely to be considerable variability in the SMBH accretion rates that may be related to the merger, such as observed in one of the dual AGN in ESO~509-IG066 \citep{kosec17}.

The inferred star formation history of M51 may also provide clues. These studies have found that the star formation rate of M51a peaked 1000--500 Myr ago at $\sim10$\msol\,yr$^{-1}$, coinciding with the second-to-last encounter. The star formation rate over the past 100 Myr, during which the most recent encounter occurred, is much lower \citep[$\sim2$\msol\,yr$^{-1}$,][]{mentuch12,eufrasio17}. If SMBH growth occurs simultaneously with star formation, this implies that the dual AGN were more active during the encounter $400-500$ Myr ago.

\subsection{The ultraluminous X-ray source population of the M51 galaxies}

The ULXs in M51 have been extensively studied in previous works, both at X-ray wavelengths \citep[e.g.][]{dewangan05,yoshida10} and longer wavelengths \citep[e.g.][]{terashima06,heida14}. Here we have presented the first systematic study of these sources at hard X-ray wavelengths, afforded by a long exposure with \nustar.  All ULXs studied so far with \nustar\ that have sufficient signal-to-noise broadband data show remarkably similar spectral shapes \citep{pintore17,koliopanos17,walton18c}, the most notable feature is a spectral turnover below 10 keV \citep[e.g.][]{bachetti13}. This feature is not seen in the X-ray spectra of any sub-Eddington accreting black holes \citep{mcclintock06}. 

The broadband ULX spectral shape is generally interpreted as the superposition of one or more disk-like components, in combination with a high energy tail. ULXs which exhibit this behaviour include the known neutron-star-accretors such as M82~X-2 \cite[albeit the spectral turnover has only been observed in the pulsed emission,][]{brightman16}, NGC~7793~P13 \citep{walton18a}, and NGC~5907~ULX1 \citep{walton18c}. For the neutron stars, the high energy tail appears to be associated with the pulsed emission and therefore with emission from the accretion column that rotates with the neutron star \citep{walton18c}. Therefore, identifying a spectral turnover in other ULXs possibly identifies them as candidate super-Eddington accretors that are potentially powered by neutron stars.

While the ULXs in M51 are faint, and the signal to noise in the \nustar\ data is not as high as the sources presented in \cite{walton18c}, we have found statistically significant evidence for a spectral turnover in three sources; ULX5, ULX7 and ULX9 at 2.6, 3 and $>3\sigma$ confidence respectively. Interestingly, only ULX8, already known to be powered by a neutron star, does not show evidence for a turn over. Nevertheless, the spectrum of ULX8 is consistent with a cut off as low as 6 keV at 90\% confidence and a turnover was identified in a \chandra\ observation when the source was observed at a higher flux \citep{brightman18}. This spectral turnover is also observed in other star-forming galaxies observed with \nustar, which are likely to be dominated by the emission from ULXs \citep{wik14,lehmer15,yukita16}. We note that a general presentation of extranuclear point sources observed by NuSTAR in nearby galaxies is presented in N. Vulic et al. 2018, ApJ, submitted.

We have also tested disk models for these sources, specifically a multi-color disk black body model with a free radial temperature profile index. While for a standard thin accretion disk this parameter is expected to be 0.75, lower values have been measured in the spectra of ULXs which are expected from slim accretion disks \citep{abramowicz88,watarai00,poutanen07}. However, for only one of our sources are the parameters of this model well constrained, for ULX5. Here the radial temperature profile is 0.6$^{+0.2}_{-0.1}$, which is consistent with either a standard accretion disk and a slim disk. For ULX3, the radial temperature index is constrained to be $>0.7$, which would rule out a slim disk scenario.

Some ULX candidates have been found to be background AGN in the past \citep[e.g.][]{gutierrez13}. We can calculate the expected number of background extragalactic sources within the area of M51 using the X-ray number counts derived from X-ray surveys. At 0.5--10 keV fluxes $>10^{-13}$ \ergcms, the X-ray source number density is 20 deg$^{-2}$ \citep{georgakakis08}. Given an area of approximately 0.02 deg$^{2}$ subtended by M51, the expected number of background sources within the galaxy is 0.4. The Poisson probability that one of the ULXs in M51 is a background source is therefore 0.27 and the probability that more than one is a background source is $\leq0.05$. However, most of the ULXs are located in the spiral arms, making their association with M51 more likely. ULX5 and ULX9 appear possibly offset from the galaxies' main structures, but for those sources we have found statistically significant evidence for a spectral turnover, so a background AGN scenario is disfavored since as stated above, these systems do not show this feature at energies below 10 keV. Finally ULX7 and ULX8 both have probable stellar counterparts, all but ensuring their position within the galaxies \citep{terashima06,earnshaw16}.

While these deep \nustar\ data have allowed us to perform some basic spectral characterization of the ULXs in M51, the count rates in the \chandra\ or \nustar\ detectors are not high enough to conduct pulsation searches. Previous discoveries have required $\sim10^{4}$ counts to detect pulsations \citep{bachetti14,fuerst16,israel17a,israel17}, whereas only a few hundred counts are observed from each source in each detector. We nevertheless carried out fast Fourier transform analyses on the lightcurves, but found no significant peaks.

 \section{Summary and Conclusions}

We have presented a broadband X-ray spectral analysis of the AGN and off-nuclear point sources in the galaxies of M51 with a simultaneous \chandra\ and deep \nustar\ observations. We have measured the intrinsic X-ray luminosities of the dual AGN with the highest fidelity yet, using the latest X-ray torus models to infer \lx\ for the Compton-thick nucleus of M51a, and resolving the nucleus of M51b for the first time with \chandra. Both SMBHs have very low accretion rates (\lamedd$<10^{-4}$) considering that the galaxies are in the process of merging. We find that the covering factor of the torus in M51a is low, which agrees with the latest results on the local AGN obscured fraction that show a low fraction at low luminosities. All of the ULXs we study show evidence for a spectral turnover, which appears to be ubiquitous when these sources are studied at high signal-to-noise.

\section{Acknowledgements} The authors thank the anonymous referee for their thorough and detailed review of our manuscript which improved it. M. Balokovi\'{c} acknowledges support from the Black Hole Initiative at Harvard University, through the grant from the John Templeton Foundation. MK acknowledges support from NASA through ADAP award NNH16CT03C. DMA acknowledges the Science and Technology Facilities Council (STFC) through grant ST/P000541/1. The work of DS was carried out at the Jet Propulsion Laboratory, California Institute of Technology, under a contract with NASA. AZ acknowledges funding from the European Research Council under the European Unions Seventh Framework Programme (FP/2007-2013) / ERC Grant Agreement n. 617001. Support for this work was provided by the National Aeronautics and Space Administration through Chandra Award Number GO7-18105X issued by the Chandra X-ray Center, which is operated by the Smithsonian Astrophysical Observatory for and on behalf of the National Aeronautics Space Administration under contract NAS8-03060. This work was also supported under NASA Contract No. NNG08FD60C, and made use of data from the {\it NuSTAR} mission, a project led by the California Institute of Technology, managed by the Jet Propulsion Laboratory, and funded by the National Aeronautics and Space Administration. We thank the {\it NuSTAR} Operations, Software and Calibration teams for support with the execution and analysis of these observations.  This research has made use of the {\it NuSTAR} Data Analysis Software (NuSTARDAS) jointly developed by the ASI Science Data Center (ASDC, Italy) and the California Institute of Technology (USA).

{\it Facilities:} \facility{\chandra\ (ACIS), \nustar, Palomar (DBSP)}

\end{document}